\documentclass[pra,twocolumn,showpacs,amsmath,amssymb]{revtex4}

\newcommand{\subsize}
{\scriptsize}

\newcommand{\sub}[1]
{_\textrm{\subsize #1}}

\newcommand{\super}[1]
{^\textrm{\subsize #1}}

\newcommand{\subtext}[1]
{\textrm{\subsize #1}}

\newcommand{\gammafktB}[2]
{\frac{\Gamma\big({#1\over2}\big)}{\Gamma\big({#2\over2}\big)}}

\newcommand{\angleint}[2]
{\int_0^{\pi/2}d#1\sin^2#1\cos^{#2}#1}

\begin{document}

\setlength\arraycolsep{1pt}

\title{Two-body correlations in $N$-body boson systems}

\author{O.~S\o rensen}
\author{D.~V.~Fedorov}
\author{A.~S.~Jensen}

\affiliation{Department of Physics and Astronomy, University of
  Aarhus, DK-8000 Aarhus C, Denmark}

\date{\today}

\begin{abstract} 
  We formulate a method to study two-body correlations in a system of
  $N$ identical bosons interacting via central two-body potentials.
  We use the adiabatic hyperspherical approach and assume a
  Faddeev-like decomposition of the wave function.  For a fixed
  hyperradius we derive variationally an optimal integro-differential
  equation for hyperangular eigenvalue and wave function. This
  equation reduces substantially by assuming the interaction range
  much smaller than the size of the $N$-body system.  At most
  one-dimensional integrals then remain.  We view a Bose-Einstein
  condensate pictorially as a structure in the landscape of the
  potential given as a function of the one-dimensional hyperradial
  coordinate.  The quantum states of the condensate can be located in
  one of the two potential minima.  We derive and discuss properties
  of the solutions and illustrate with numerical results.  The
  correlations lower the interaction energy substantially.  The new
  multi-body Efimov states are solutions independent of details of the
  two-body potential.  We compare with mean-field results and
  available experimental data.
\end{abstract}


\maketitle

\section{Introduction}

The average properties of an $N$-body system are often investigated in
the mean-field approximation \cite{bay96,dal99,pet01}, where the wave
function is a product of one-particle amplitudes.  This excludes a
priori effects of particle correlations, which often are responsible
for decisive features or phenomena.  A prominent example is the use of
the bare nucleon-nucleon interaction in computations of nuclear ground
state structure.  The Hartree-Fock results are catastrophic, either
producing unbound systems or collapsed point-like structures with
infinite binding energy \cite{sie87}.  The correlations must be dealt
with here.  One way is to use effective interactions and maintain the
same Hilbert space of independent particles \cite{bra85}.  One can get
a long way by this procedure, but the particles are still not
correlated in the wave functions, although reasonable sizes and
binding energies are found \cite{bra85,sie87,dal99}.  Any effect of
correlations can therefore not be tested experimentally except as
deviations from the mean-field results.

Another example, where correlations are essential, is the decay of
Bose-Einstein condensates \cite{nie99,esr99}, which cannot be referred
to as the lowest energy solution.  Indeed, the $N$-body system has
many lower lying states as immediately realized from the fact that two
and more atoms form bound states.  The energy is then already lower
than the energy of the condensate, which therefore eventually decays
into these lower lying structures.  The decay can be either three-body
recombination possibly enhanced by the presence of other particles
\cite{nie99,esr99,bra02b}, or macroscopic collapse into a structure of
very small spatial dimension, which subsequently then recombine into
more favorable bound states \cite{pit96,adh02b,don01}.

The direct recombination process is increasingly probable with larger
scattering length \cite{nie99,esr99}, and also the macroscopic
collapse must increase strongly with scattering length \cite{sor02}.
The limit of infinite scattering length is traditionally considered as
difficult to solve as exemplified by the three-body system where the
delicate Efimov states could occur \cite{efi70,fed93}.  The structure
of the $N$-body system in this limit is at least as difficult as the
three-body problem \cite{nie01}, but the growing interest
\cite{lie98,lie01,blu01,cow01} and the difficulties in this area
demand new approaches.

The macroscopic collapse is conceptually very similar to the nuclear
fission process where the (liquid) nucleus in a collective process is
divided into two or more pieces.  One difference is that in fission
the fragments move away from each other, whereas (gaseous)
Bose-Einstein condensates first collapse into a more dense state and
after recombination into smaller subsystems the fragments move apart
\cite{don01}.  These collapse mechanisms can to a large extent be
described in a mean-field picture \cite{boh98,adh02b}, but especially
for condensates the correlations could be very important first to
establish the collective coordinate and the corresponding potential
energy and second by influencing the collapse process itself
\cite{sor01,sor02}.

The essential ingredients in a description of correlations in $N$-body
systems are the techniques used to solve few-body problems
\cite{car98,nie01}.  Isolated two-body systems are easily solved and
the key is very likely in handling of the three-body problem.  This
expectation arises since two particles plus all remaining particles
effectively is a three-body problem, and this type of two-body
correlations beyond the mean-field inside the $N$-body system is
probably dominating.  This is also the philosophy in the Faddeev
\cite{fad61} and Yakubovsky \cite{yak67} equations, where the two-body
amplitudes eventually are the basic quantities.  The Yakubovsky
reduction is rigorous, but very cumbersome in its full glory.
However, such formulations may provide inspiration to practical
approximations.

Unfortunately, it is impossible to include correlations directly on
top of mean-field calculations with attractive zero range interactions
\cite{phi97,fed01}.  The Gross-Pitaevskii equation without a confining
external trap can only be used for repulsion in a product wave
function.  The lowest state for an attraction would correspond to a
divergent collapsed non-physical solution, but an additional external
field is able to hold metastable solutions at larger distances.
Skyrme Hartree-Fock is more sophisticated and an attraction leading to
a bound system does avoid collapsed solutions \cite{bra85}.  Including
correlations triggers the Thomas effect \cite{tho35}, where collapsed
three-body states would appear inside the many-body system
\cite{fed01}.  Renormalization can be invoked to cure these
divergences and maintain the simplicity of the zero-range interaction
\cite{fed01b,ols01}, but correlations are still not included.
Obviously a finite range interaction would also prevent the disaster
when correlations are allowed.

To study correlations the form of the wave function must be flexible
enough to include the corresponding degrees of freedom.  As we learned
from the mean-field approximation sizes and binding energies may be
rather accurate even with an imprecise wave function
\cite{neg82,sor01}.  Thus for a specific purpose it seems possible to
design a relatively simple wave function, which for other purposes may
be a rather poor approximation, but precisely accounting for the
desired degrees of freedom.  A promising form of a correlated wave
function suggested for nucleons \cite{rip84} was recently extended to
more general systems \cite{bar99a}.  The simplest structure is clearly
found in Bose-Einstein condensates.  Here was recently introduced a
formulation with generalized hyperspherical coordinates and an
adiabatic expansion with the hyperradius as the adiabatic coordinate
\cite{boh98}.  Only the crude approximations of a zero-range
interaction and the lowest (constant and thus non-correlated)
hyperspherical angular wave function were used in \cite{boh98}.

In this paper we shall use the simplifications of the completely
symmetric wave function.  These systems reveal features only
explainable by correlations and we can then already here obtain
results of practical interest.  The purpose of this paper is to
formulate the details of a recently suggested \cite{sor01} method to
include two-body correlations in an $N$-body system of identical
bosons.  We shall describe the qualitative features of the key
quantities, give numerical examples, compare with established methods
and available experimental data and provide predictions going beyond
the present knowledge.  The model is apparently rich with many
unexplored possibilities perhaps especially by application to systems
with large scattering length, to the dynamical evolution, and to other
systems of different symmetries.

\section{The $N$-body problem and the hyperspherical formulation}

\label{sec:n-body-problem}

A formulation accounting for particle correlations in an $N$-body
system must by definition go beyond the mean-field approximation. The
Faddeev-Yakubovsky equations could be an appropriate starting point.
However, first the form of the Hamiltonian describing the system must
be decided.  Then a convenient set of coordinates must be chosen in
harmony with the formulation and the anticipated approximations.
Derivations of suitable equations of motion are then possible and
their properties can be investigated. This section describes how we
choose two-body interactions, hyperspherical coordinates, a
Faddeev-like decomposition of the wave function, use an adiabatic
expansion, and assume at first only s-waves.

\subsection{The hyperspherical coordinates}

The system of $N$ identical interacting bosons of mass $m$ may be
described by $N$ coordinate vectors $\vec{r}_{i}$ and momenta
$\vec{p}_{i}$, labeling the particles by the index $i=1,\ldots,N$.
Here a more suitable choice of coordinates is the center of mass
coordinates $\vec R=\sum_{i=1}^N\vec r_i/N$, the $N-1$ relative Jacobi
vectors $\vec{\eta}_k$ with $k=1,2,\ldots,N-1$ \cite{bar99a,bar99b}
\begin{eqnarray}
  \label{e2}
  \vec{\eta}_k
  &=&
  \sqrt{\frac{N-k}{N-k+1}}
  \biggl(
  \vec{r}_{N-k+1}-
  \frac1{N-k}\sum_{j=1}^{N-k}\vec{r}_j
  \biggr)
  \;,
\end{eqnarray}
and their associated momenta. These Jacobi coordinates are illustrated
for up to six particles in fig.~\ref{fig:jacobi.trees}a. We use the
notation $\eta_k\equiv|\vec\eta_k|$, so $\eta_{N-1}$ is proportional
to the distance between particles $1$ and $2$, $\eta_{N-2}$ is
proportional to the distance between particle $3$ and the center of
mass of $1$ and $2$, $\eta_{N-3}$ is proportional to the distance
between particle $4$ and the center of mass of the first three
particles, etc.

Hyperspherical coordinates are now defined in relation to the Jacobi
vectors.  One length, the hyperradius $\rho$, is defined by
\begin{eqnarray}
  \label{e4}
  \rho_l^2
  \equiv
  \sum_{k=1}^l\eta_k^2
  \;,\quad
  \rho^2
  \equiv
  \rho_{N-1}^2
  =
  \frac{1}{N} \sum_{i<j}^N r_{ij}^2
  \;,
\end{eqnarray}
where $r_{ij} \equiv |\vec{r}_{i}-\vec{r}_{j}|$.  The $N-2$
hyperangles $\alpha_k\in[0,\pi/2]$ for $k=2,3,\ldots,N-1$ relate the
length of the Jacobi vectors to the hyperradius via the definition
\begin{eqnarray}
  \label{e3}
  \sin\alpha_k
  \equiv
  \frac{\eta_k}{\rho_k}
  \;.
\end{eqnarray}
Since $\rho_1=\eta_1$ the variable $\alpha_1=\pi/2$ is superfluous,
but is for convenience often included in the notation.  Remaining are
the $2(N-1)$ angles $\Omega_\eta^{(k)}= (\theta_k, \varphi_k)$ for
$k=1,2,\ldots,N-1$ defining the directions of the $N-1$
$\vec\eta_k$-vectors, i.e.~$\theta_k\in[0,\pi]$ and $\varphi_k \in
[0,2\pi]$.  All angles are collectively denoted by $\Omega \equiv
\{\alpha_k,\theta_k,\varphi_k\},\; k=1,2,\ldots,N-1$.  In total
$\Omega$ and $\rho$ amount to $3(N-1)$ degrees of freedom and the
center of mass coordinates $\vec{R}$ amount to three.  These
coordinates are connected by
\begin{eqnarray}
  \label{e5}
  \sum_{i=1}^Nr_i^2=
  \frac1N\sum_{i<j}^Nr_{ij}^2
  +\frac1N\bigg(\sum_{i=1}^N\vec{r}_i\bigg)^2
  =\rho^2+NR^2 \; .
\end{eqnarray}

The total volume element is $\prod_{i=1}^N d^3\vec r_i =
N^{3/2}d^3\vec R\prod_{k=1}^{N-1}d^3\vec\eta_k$, where the part
depending on relative coordinates is $\prod_{k=1}^{N-1}d^3\vec\eta_k$.
In hyperspherical coordinates this relative part becomes
\begin{eqnarray}
  \label{e6}
  &&\prod_{k=1}^{N-1}d^3\vec\eta_k
  =d\rho \rho^{3N-4} 
  \; d\Omega_{N-1}
  \;,
  \\
  \label{e7}
  &&d\Omega_k
  =
  d\Omega_\alpha^{(k)}
  \;
  d\Omega_\eta^{(k)}
  \;
  d\Omega_{k-1}
  \;,
  \\
  \label{e8}
  &&d\Omega_\alpha^{(k)}
  =d\alpha_k\sin^2\alpha_k\cos^{3k-4}\alpha_k
  \;,
\end{eqnarray}
where $d\Omega_\eta^{(k)} =d\theta_k\sin\theta_kd\varphi_k$ is the
familiar angular volume element in spherical coordinates.  The
recursion stops at $d\Omega_1=d\Omega_\eta^{(1)}$.  Since the angle
$\alpha_{N-1}$ is related directly to a two-body distance, $r_{12}$,
by $\sin\alpha_{N-1}=\eta_{N-1}/\rho_{N-1}=r_{12}/(\sqrt2\rho)$, the
volume element in eq.~(\ref{e8}) related to this angle is especially
important: $d\Omega_\alpha^{(N-1)} = d\alpha_{N-1} \sin^2\alpha_{N-1}
\cos^{3N-7}\alpha_{N-1}$.

The angular volume integrals can be computed \cite{sch68}, i.e.~$\int
d\Omega_\alpha^{(k)} = \sqrt\pi\Gamma(3(k-1)/2)/(4\Gamma(3k/2))$ and $\int
d\Omega_\eta^{(k)} = 4\pi$, where $\Gamma(x)$ is the gamma function.

Using eq.~(\ref{e6}) an angular matrix element of an operator
$\hat{O}$ for fixed $\rho$ is then
\begin{eqnarray}
  \label{e9}
  \langle\Psi|\hat{O}|\Phi\rangle_\Omega
  =
  \int d\Omega_{N-1}\;\Psi^*(\rho,\Omega)\;\hat{O}\;\Phi(\rho,\Omega)
  \;,
\end{eqnarray}
which in general is a function of $\rho$.

\subsection{Hamiltonian}

We consider $N$ identical particles of mass $m$ interacting through
short range two-body potentials. For completeness we will throughout
the paper include an external trap confining the particles to a
limited region of space through a simple three-dimensional harmonic
oscillator potential $m\omega^2r_i^2/2$. The trap is relevant for
Bose-Einstein condensation but can easily be omitted.  With a two-body
central potential $V_{ij}=V(r_{ij})$ the total Hamiltonian is given by
\begin{eqnarray}
  \hat H\sub{total}
  =
  \sum_{i=1}^N
  \bigg(
  \frac{\hat p_i^2}{2m}+\frac12m\omega^2r_i^2
  \bigg)
  +\sum_{i<j}^N V(r_{ij})
  \;,\quad
\end{eqnarray}
which, using eq.~(\ref{e5}), is separable into a part only involving
the center of mass coordinates and a part only involving relative
coordinates.  We can see this by subtracting
\begin{eqnarray}
  \label{eq:1}
  \hat H\sub{cm}
  \equiv
  \frac{\hat P_{R}^2}{2M}
  +\frac12M\omega^2R^2
  \;,
\end{eqnarray}
where $\vec P_{R} \equiv \sum_i \vec p_i$ is the total momentum and
$M=Nm$ is the total mass of the system, i.e.
\begin{eqnarray}
  \hat H
  \equiv
  \hat H\sub{total}-\hat H\sub{cm}
  =\sum_{i=1}^N\frac{\hat p_i^2}{2m}-\frac{\hat P_R^2}{2M}+
  \nonumber\\
  \sum_i^N\frac12m\omega^2r_i^2
  -\frac12M\omega^2R^2
  +\sum_{i<j}^NV_{ij}
  \;.
\end{eqnarray}
Using eq.~(\ref{e5}) and denoting the intrinsic kinetic energy
operator by $\hat T \equiv
\sum_{i=1}^N\hat p_i^2/(2m)-\hat P_R^2/(2M)$ we can write
\begin{eqnarray}
  \hat H=
  \hat T
  +\frac12m\omega^2\rho^2
  +\sum_{i<j}^NV_{ij}
  \;.
\end{eqnarray}
In hyperspherical coordinates $\hat T$ can be rewritten as
\cite{bar99a,bar99b}
\begin{eqnarray}
  \label{e14}
  \hat T
  =
  -\frac{\hbar^2}{2m}
  \biggl[\frac{1}{\rho^{3N-4}}\frac{\partial}{\partial\rho}
  \rho^{3N-4} \frac{\partial}{\partial\rho}
  -\frac{\hat\Lambda_{N-1}^2}{\rho^2}
  \biggr]
  \;,
\end{eqnarray}
with the dimensionless angular kinetic energy operator
$\hat\Lambda_{N-1}^2$ recursively defined by
\begin{eqnarray}
  \label{e18}
  &&
  \hat\Lambda_k^2
  =\hat\Pi_k^2+
  \frac{\hat\Lambda_{k-1}^2}{\cos^2\alpha_k}
  +\frac{\hat l_k^2}{\sin^2\alpha_k}
  \;,\quad
  \\  \label{e18a}
  &&
  \hat\Pi_k^2
  =-\frac{\partial^2}{\partial\alpha_k^2}+
  \frac{3k-6-(3k-2)\cos 2\alpha_k}{\sin 2 \alpha_k}
  \frac{\partial}{\partial\alpha_k}
  \;,\quad
\end{eqnarray}
where $\hbar\hat l_k$ is the angular momentum operator associated with
$\vec\eta_k$.  The recursion stops at $\hat\Lambda_1^2=\hat l_1^2$.
The angular kinetic energy operator is thus a sum of derivatives with
respect to the various hyperspherical angles.  Convenient
transformations to get rid of first derivatives in eqs.~(\ref{e14})
and (\ref{e18a}) are
\begin{eqnarray}
  \label{e22}
  &&
  -\frac{2m}{\hbar^2}\hat T_\rho
  \equiv
  \rho^{-(3N-4)}
  \frac{\partial}{\partial\rho} \rho^{3N-4} \frac{\partial}{\partial\rho} 
  =
  \\
  &&
  \quad
  \rho^{-(3N-4)/2}
  \bigg(
  \frac{\partial^2}{\partial\rho^2}-\frac{(3N-4)(3N-6)}{4\rho^2}
  \bigg)
  \rho^{(3N-4)/2}
  \;,\;
  \nonumber\\
  &&
  \hat\Pi_k^2
  =\sin^{-1}\alpha_k\cos^{-(3k-4)/2}\alpha_k
  \bigg(-\frac{\partial^2}{\partial\alpha_k^2}-\frac{9k-10}2+
  \nonumber\\
  &&
  \quad
  \frac{(3k-4)(3k-6)}{4}\tan^2\alpha_k
  \bigg)
  \sin\alpha_k\cos^{(3k-4)/2}\alpha_k
  \;.\;
  \label{e19} 
\end{eqnarray}

The Hamiltonian $\hat H$ can now be rewritten as
\begin{eqnarray}
  \label{e20}
  \hat H
  &=&
  \hat T_\rho
  +\frac12m\omega^2\rho^2
  +\frac{\hbar^2}{2m\rho^2}
  \hat h_\Omega
  \;,
  \\
  \label{eq:22}
  \hat h_\Omega
  &\equiv&
  \hat\Lambda^2_{N-1}
  +\sum_{i<j}^Nv_{ij}
  \;,\qquad
\end{eqnarray}
where $v_{ij} \equiv 2m\rho^2V_{ij}/\hbar^2$ is a dimensionless
potential, $\hat T_\rho$ is the radial kinetic energy operator, and
$\hat h_\Omega$ is the dimensionless angular Hamiltonian.  The
intrinsic Hamiltonian, $\hat H$, thus contains a part depending on
$\rho$ and a part, $\hat h_\Omega$, depending on both $\rho$
(parametrically) and $\Omega$.

\subsection{Equations of motion}

Since the total Hamiltonian is given as $\hat H\sub{total} = \hat
H\sub{cm}+\hat H$, the total wave function for the $N$-particle system
can without loss of generality be written as a product of a function,
$\Upsilon$, depending only on $\vec R$ and a function, $\Psi$,
depending on $\rho$ and the $3N-4$ angular degrees of freedom
collected in $\Omega$: $\Upsilon(\vec R) \Psi(\rho,\Omega)$.  The
center of mass motion for the total mass $M=Nm$ is given by
\begin{eqnarray}
  \hat H\sub{cm}\Upsilon(\vec R)=E\sub{cm}\Upsilon(\vec R)
  \;,
\end{eqnarray}
and as seen from eq.~(\ref{eq:1}) the energy spectrum is that of a
harmonic oscillator, i.e.~$E_{\subtext{cm},n} = \hbar\omega(n+3/2)$,
where $n$ is a non-negative integer.

The relative wave function $\Psi(\rho,\Omega)$, obeying the stationary
Schr\"odinger equation,
\begin{eqnarray}
  \label{eq:equations-motion}
  \hat H\Psi(\rho,\Omega)
  =
  E\Psi(\rho,\Omega)
  \;,
\end{eqnarray}
is for each value of the hyperradius $\rho$ expanded as
\begin{eqnarray}
  \label{e24}
  \Psi(\rho,\Omega)
  =
  \rho^{-(3N-4)/2}\sum_{n=0}^\infty f_n(\rho)\Phi_n(\rho,\Omega)
  \;,
\end{eqnarray}
where the factor $\rho^{-(3N-4)/2}$ is included to eliminate first
derivatives in $\rho$, see eq.~(\ref{e22}).  Here the hyperradial wave
functions, $f_n(\rho)$, are the expansion coefficients for fixed
$\rho$ on the complete set of solutions $\Phi_n(\rho,\Omega)$ obtained
by solving the angular eigenvalue equation:
\begin{eqnarray}
  \label{e26}
  (\hat h_\Omega-\lambda_n)\Phi_n(\rho,\Omega)
  =0
  \;,
\end{eqnarray}
where $\lambda_n$ is the angular eigenvalue, which depends on $\rho$.
We will usually apply the normalization
$\langle\Phi_n|\Phi_m\rangle_\Omega = \delta_{n,m}$.

In complete analogy to the technique employed for $N=3$ \cite{jen97}
we insert eq.~(\ref{e24}) in eq.~(\ref{eq:equations-motion}), use
eqs.~(\ref{e20}) and (\ref{e26}), and finally project the resulting
equation onto the angular eigenfunctions $\Phi_n(\rho,\Omega)$. We
then arrive at a set of coupled radial equations
\begin{widetext}
\begin{eqnarray}
  \label{eq:coupled.radial.equation}
  \bigg(
  -\frac{d^2}{d\rho^2}-\frac{2mE}{\hbar^2}+
  \frac{\lambda_n(\rho)}{\rho^2}+
  \frac{(3N-4)(3N-6)}{4\rho^2}+
  \frac{\rho^2}{b\sub t^4}
  -Q_{nn}^{(2)}(\rho)
  \bigg)
  f_n(\rho)
  = 
  \sum_{n'\ne n}\bigg(
  2Q_{nn'}^{(1)}(\rho)\frac{d}{d\rho}+Q_{nn'}^{(2)}(\rho)
  \bigg)
  f_{n'}(\rho)
  \;,\quad
\end{eqnarray}
\end{widetext}
where the trap length is $b\sub t\equiv\sqrt{\hbar/(m\omega)}$ and the
coupling terms $Q_{nn'}^{(i)}$ ($Q_{nn}^{(1)}=0$) are defined as
\begin{eqnarray}
  \label{eq:coupling.terms}
  Q_{nn'}^{(i)}(\rho)
  \equiv
  \frac
  {\big\langle
    \Phi_n(\rho,\Omega)
    \big|
    \big[\frac{\partial}{\partial\rho}\big]^i
    \big|
    \Phi_{n'}(\rho,\Omega)
    \big\rangle_\Omega}
  {\big\langle
    \Phi_n(\rho,\Omega)
    \big|
    \Phi_n(\rho,\Omega)
    \big\rangle_\Omega}
  \;.\qquad
\end{eqnarray}
The angular eigenvalues $\lambda_n$ enter these coupled equations as a
radial potential. The total diagonal effective radial potential,
$U_n(\rho)$, entering on the left hand side of
eq.~(\ref{eq:coupled.radial.equation}) is:
\begin{eqnarray}
  \label{eq:radial.potential}
  \frac{2mU_n}{\hbar^2}
  \equiv
  \frac{\lambda_n}{\rho^2}+
  \frac{(3N-4)(3N-6)}{4\rho^2}+
  \frac{\rho^2}{b\sub t^4}
  -Q_{nn}^{(2)}
  \;.
\end{eqnarray}
This includes a $\rho^2$-term due to the external harmonic field, a
$\rho^{-2}$ centrifugal barrier-term due to the transformation of the
radial kinetic energy operator, the angular potential $\lambda_n$, and
the diagonal coupling term $Q_{nn}^{(2)}$. 

If the non-diagonal coupling terms are neglected, i.e.~the right hand
side of eq.~(\ref{eq:coupled.radial.equation}) vanishes, the equations
simplify significantly to
\begin{eqnarray}
  \label{e32}
  \bigg(
  -\frac{\hbar^2}{2m}
  \frac{d^2}{d\rho^2}
  +U_n(\rho)-E_{n,q}\bigg)
  f_{n,q}(\rho)=0
  \;.
\end{eqnarray}
We have here included another index $q$ to indicate a spectrum of
energies and radial wave functions obtained for each angular solution
$n$.

\subsection{The angular eigenvalue equation}
\label{sec:angular-equation}

The eigenvalue $\lambda$ from eq.~(\ref{e26}) is the key quantity
carrying essentially all information about the two-body interactions
and therefore about possible correlations as well. The technique and
approximations used to find $\lambda$ are then especially important.

\subsubsection{Decomposition of the angular wave function}

The angular eigenvalue is obtained by solving eq.~(\ref{e26}). For
each $\rho$ we first assume a decomposition of the wave function
$\Phi$ (omitting the index $n$) in additive components $\Phi_{ij}$,
i.e.
\begin{eqnarray}
  \label{e34}
  \Phi(\rho,\Omega)
  =
  \sum_{i<j}^N\Phi_{ij}(\rho,\Omega)
  \;,
\end{eqnarray}
where each term $\Phi_{ij}$ is a function of $\rho$ and all angular
coordinates $\Omega$.  This decomposition is in principle exact, since
each term in itself is sufficient when all $\Omega$ degrees of freedom
are allowed.  At first this ansatz seems clumsy by introducing an
overcomplete basis.  However, the indices $i$ and $j$ indicate special
emphasis on the particle pair $i-j$. The component $\Phi_{ij}$ is
expected to carry the information associated with two-body
correlations of this particular pair. This has no significance before
it is exploited in numerical techniques or subsequent approximations.

Rewriting the wave function obeying the Schr\"odinger equation as a
sum of terms has been very successful in three-body computations. The
advantage is that the correct boundary conditions are simpler to
incorporate as expressed in the original formulation by Faddeev
\cite{fad61} intended for scattering.  Still mathematically nothing is
gained or lost in this Faddeev-type of decomposition.  For very weakly
bound and spatially very extended three-body systems, s-waves in each
of the Faddeev components are sufficient to describe the system. This
is exceedingly pronounced for large scattering lengths where the
delicate Efimov states appear \cite{jen97,nie01}.

The present $N$-body problem is of course in general more complicated.
However, for dilute condensates essential similarities remain,
i.e.~the relative motion of two particles on average far from each
other is most likely dominated by s-wave contributions. Each particle
cannot detect any directional preference arising from higher partial
waves.  Only the monopole prevails. Implementing these ideas in the
present context imply that each amplitude $\Phi_{ij}$ for a fixed
$\rho$ only should depend on the distance $r_{ij}$ between the two
particles.  For that purpose we define a two-index parameter
$\alpha_{ij}$ by
\begin{eqnarray}
  \label{eq:8}
  \sin\alpha_{ij}
  \equiv
  \frac{r_{ij}}{\sqrt2\rho}
  \;,
\end{eqnarray}
which is distinctively different from the $\alpha_k$'s of
eq.~(\ref{e3}).  Thus we assume
\begin{eqnarray}
  \Phi_{ij}(\rho,\Omega)
  \simeq
  \phi_{ij}(\rho,\alpha_{ij})
  \;.
\end{eqnarray}

The boson symmetry implies that all the functions $\phi_{ij}$ are
equal and that we should not distinguish, so we therefore omit the
indices.  We then arrive at the angular wave function
\begin{eqnarray} \label{e77}
  \Phi(\rho,\Omega)
  =
  \sum_{i<j}^N\phi(\rho,\alpha_{ij}) = \sum_{i<j}^N\phi(\alpha_{ij}) 
  \;,
\end{eqnarray}
where we used $\phi_{ij}(\rho,\alpha_{ij}) = \phi(\rho,\alpha_{ij})
\equiv \phi(\alpha_{ij})$ with omission of the coordinate $\rho$ in
the last notation.  The wave function in eq.~(\ref{e77}) is symmetric
with respect to interchange of two particles, $i\leftrightarrow j$,
since $\alpha_{ij}=\alpha_{ji}$ and since terms like
$\phi(\alpha_{ik})+\phi(\alpha_{jk})$ always appear symmetrically.

This ansatz of only s-waves dramatically simplifies the angular wave
function.  The original overcomplete Hilbert space is now reduced, so
not every angular wave function can be expressed in this remaining
basis.  Thus rigorously the reduction resulted in an incomplete basis,
but the degrees of freedom remaining in eq.~(\ref{e77}) are expected
to be precisely those needed to describe the main features of the
condensate. The approximations are tailored to the problem under
investigation.

\subsubsection{Faddeev-like equations}

Inserting the ansatz of eq.~(\ref{e77}) along with eq.~(\ref{eq:22})
in eq.~(\ref{e26}) yields the angular equation
\begin{eqnarray}
  \label{eq:10}
  \Big(
  \hat\Lambda_{N-1}^2+\sum_{k<l}^Nv_{kl}-\lambda
  \Big)
  \sum_{i<j}^N\phi_{ij}
  =
  0  
  \;,
\end{eqnarray}
with $\phi_{ij}=\phi(\alpha_{ij})$.  Rearrangement of summations leads
to
\begin{eqnarray}\label{eq:11}
  \sum_{i<j}^N\bigg[
  \Big(\hat\Lambda_{N-1}^2-\lambda\Big)\phi_{ij}
  +v_{ij}\sum_{k<l}^N\phi_{kl}
  \bigg]
  =0 \;.
\end{eqnarray}
For three particles the Faddeev equations are obtained by assuming
that each term in the square brackets separately is zero \cite{jen97}.
The same assumption for the $N$-particle system results in the
$N(N-1)/2$ Faddeev-like equations
\begin{eqnarray}\label{eq:angular-equation1}
  \Big(\hat\Lambda_{N-1}^2-\lambda\Big)\phi_{ij}
  +v_{ij}\sum_{k<l}^N\phi_{kl}
  =
  0
  \;,
\end{eqnarray}
which are actually identical due to symmetry.  We shall not in the
present paper rely on the validity of this assumption, but only use it
to illustrate the procedure in the general discussion.

Choosing $i=1$ and $j=2$, with the ansatz for the wave function,
eq.~(\ref{e77}), the kinetic energy operator $\hat\Lambda^2_{N-1}$ in
eq.~(\ref{e18}) reduce to $\hat\Pi^2_{N-1}$, because
${\hat\Lambda}_{N-2}^2 \phi_{12} = 0$ and $\hat l_{N-1}^2 \phi_{12}
=0$.  Since $\vec \eta_{N-1} = (\vec r_2 -\vec r_1)/\sqrt{2}$ and
$\rho_{N-1} = \rho$ we have $\alpha_{N-1} = \alpha_{12}$ (compare
eqs.~(\ref{e3}) and (\ref{eq:8})), so only derivatives with respect to
$\alpha_{12}$ remains.  Thus it is convenient to introduce the
notation $\hat\Pi^2_{12} \equiv \hat\Pi^2_{N-1}$.

In the sum over angular wave function components in
eq.~(\ref{eq:angular-equation1}) only three different types of terms
appear. Assuming $i=1$ and $j=2$ these types are classified by the set
$\{k,l\}$ either having two, one, or zero numbers coinciding with the
set $\{1,2\}$.  Then eq.~(\ref{eq:angular-equation1}) is rewritten as
\begin{eqnarray}
  &&
  0=
  \Big(\hat\Pi^2_{12}+v(\alpha_{12})-\lambda\Big)\phi(\alpha_{12})+
  \label{e73} \\
  &&
  v(\alpha_{12})\Big(
  \sum_{l=3}^N\phi(\alpha_{1l})+
  \sum_{l=3}^N\phi(\alpha_{2l})+
  \sum_{k\ge 3,l>k}^N\phi(\alpha_{kl})
  \Big)
  \;,
  \nonumber
\end{eqnarray}
with $v(\alpha_{ij}) = 2m\rho^2V(\sqrt2\rho\sin\alpha_{ij})/\hbar^2$.
Multiplying this equation from the left by $\phi(\alpha_{12})$,
followed by integration over all angular space except $\alpha_{12}$
results in an integro-differential equation in $\alpha \equiv
\alpha_{12}$ of the form
\begin{eqnarray}
  \label{eq:angular-equation3}
  &&
  \Big(\hat\Pi^2_{12}+v(\alpha)-\lambda\Big)\phi(\alpha)+
  v(\alpha)
  2(N-2)\int d\tau\;\phi(\alpha_{13})
  \nonumber\\
  &&
  +v(\alpha)\frac12(N-2)(N-3)\int d\tau\;\phi(\alpha_{34})
  =0
  \;.
\end{eqnarray}
Here $d\tau\propto d\Omega_{N-2}$ is the angular volume element
excluding the $\alpha$-dependence; the normalization is $\int
d\tau=1$.  This projection leaves for every value of $\alpha$ only two
different integrals due to symmetry between the first and second sum
in eq.~(\ref{e73}).  Both of the remaining integrals can analytically
be reduced to one dimension. The results, collected in appendix
\ref{sec:app.fadd-like-equat}, are denoted by
\begin{eqnarray}
  \label{eq:intf34}
  \int d\tau\; \phi(\alpha_{34})
  =
  \hat R_{34}^{(N-2)} \phi(\alpha)
  \;,
  \\
  \label{eq:intf13}
  \int d\tau\;\phi(\alpha_{13})
  =
  \hat R_{13}^{(N-2)} \phi(\alpha)
  \;,
\end{eqnarray}
where $\hat R_{ij}^{(N-2)}$ is an operator acting on the function
$\phi(\alpha)$ resulting in a new function of $\alpha$.
Mathematically $\hat R$ resembles a rotation operator, hence the
choice of notation.  Eq.~(\ref{eq:angular-equation3}) can now be
written as
\begin{eqnarray}
  \label{eq:35}
  0
  &=&
  \bigg(
  \hat\Pi^2_{12}+v(\alpha)-\lambda+
  2(N-2)v(\alpha)\hat R_{13}^{(N-2)}
  \nonumber\\
  &&+
  \frac12(N-2)(N-3)v(\alpha)\hat R_{34}^{(N-2)}
  \bigg)
  \phi(\alpha)
  \;,
\end{eqnarray}
which is linear in the function $\phi$.

\subsubsection{Angular kinetic energy eigenfunctions}

We first consider eq.~(\ref{eq:35}) for non-interacting particles,
i.e.~$v=0$.  Using the transformation in eq.~(\ref{e19}) to get rid of
first derivatives we find
\begin{eqnarray}
  \label{eq:2}
  &&
  \Big(-\frac{d^2}{d\alpha^2}+
  \frac{(3N-7)(3N-9)}{4}\tan^2\alpha
  \nonumber\\
  &&
  -\frac{9N-19}{2}
  -\lambda\Big)
  \tilde\phi(\alpha)
  =
  0
  \;,
\end{eqnarray}
where $\tilde\phi(\alpha)$ is defined as the reduced angular wave
function
\begin{eqnarray}
  \label{eq:angular-equation-1}
  \tilde\phi(\alpha)
  \equiv
  \sin\alpha\cos^{(3N-7)/2}\alpha
  \;\phi(\alpha)
  \;,
\end{eqnarray}
in perfect analogy to the transformation from radial to reduced
radial wave function for the two-body problem.
Since $\phi(\alpha)$ for a physical state cannot diverge at $\alpha=0$
or $\alpha=\pi/2$, the boundary condition for the reduced
angular wave function is $\tilde\phi(0)=\tilde\phi(\pi/2)=0$.

The (non-reduced) solutions to eq.~(\ref{eq:2}) is given by the Jacobi
polynomials \cite{abr}
\begin{eqnarray}
  \label{eq:3}
 \phi_K(\alpha)=
  P_\nu^{(1/2,(3N-8)/2)}(\cos2\alpha)
  \;,
\end{eqnarray} 
where the hyperspherical quantum number $K=2\nu=0,2,4,\ldots$ denotes
the angular kinetic energy eigenfunction with $\nu$ nodes.  The
corresponding angular eigenvalues are $\lambda_K=K(K+3N-5)$.  The
lowest eigenvalue is zero corresponding to a constant eigenfunction 
$P_0^{(1/2,(3N-8)/2)}=1$.

\begin{figure}[htbp]
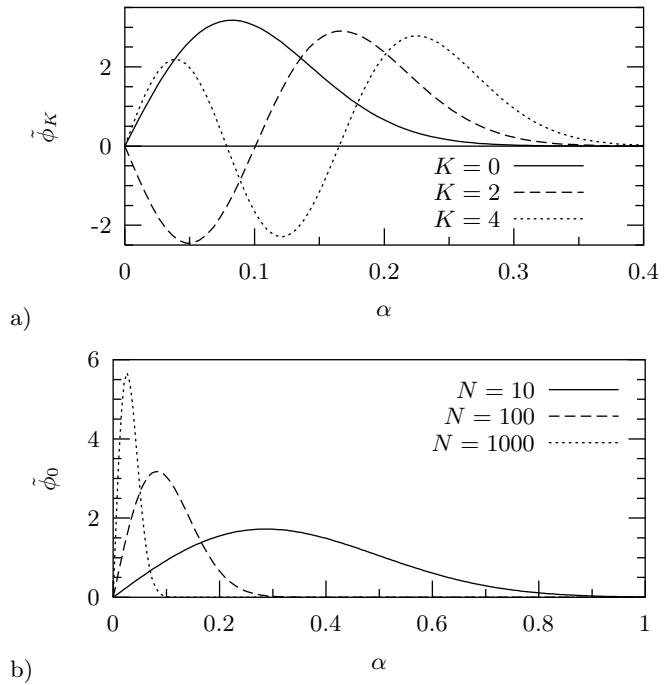

  \begin{center}
    %a)\input{/usr/users/oles/Few-body/Skriblerier/DelA/JacobyP/jacobyp_N100.tex}
    a)\input{fig3a.tex}
    \hspace{.5cm}
    %b)\input{/usr/users/oles/Few-body/Skriblerier/DelA/JacobyP/jacobyp_N.tex}
    b)\input{fig3b.tex}
  \end{center} 
  \vspace{-.5cm}
  \caption
  [] {The reduced angular wave function $\tilde\phi_K$ defined in
    eqs.~(\ref{eq:angular-equation-1}) and (\ref{eq:3}), for a)
    $N=100$ and $K=0,2,4$ and b) $K=0$ and $N=10,100,1000$.  The
    normalization is $\int_0^{\pi/2} d\alpha\;|\tilde\phi_K(\alpha)|^2
    = 1$.}
  \label{fig13}
  \vspace{-.2cm}
\end{figure}

In fig.~\ref{fig13}a are shown examples of the reduced angular kinetic
energy eigenfunctions for $N=100$ and the lowest three eigenvalues.
The constant wave function, $\phi_{K=0}$, is in the figure represented
by $\tilde\phi_0(\alpha) = \sin\alpha\;\cos^{(3N-7)/2}\alpha$, where
$|\tilde\phi_0|^2$ then simply is the volume element in
$\alpha$-space.  The oscillations are located at relatively small
$\alpha$-values. As seen in fig.~\ref{fig13}b their amplitudes
decrease as $1/\sqrt N$ due to the centrifugal barrier proportional to
$\tan^2\alpha$ in eq.~(\ref{eq:2}). Thus, as $N$ increases the
probability becomes increasingly concentrated in a smaller and smaller
region of $\alpha$-space around $\alpha =0$.

Some solutions may be spurious, i.e.~each component $\phi$ is
non-vanishing but the full wave function $\Phi$ in eq.~(\ref{e77}) is
identically zero. From eqs.~(\ref{eq:10}) and (\ref{eq:11}) we see
that the component $\phi$ for a spurious solution is an eigenfunction
of the angular kinetic energy operator. This special situation occurs
for the $K=2$ eigenfunction in eq.~(\ref{eq:3}), which satisfies
\begin{eqnarray}
  \label{eq:K2spurious}
  \int d\tau\;\sum_{i<j}^N\phi_{K=2}(\alpha_{ij})=0\;.
\end{eqnarray}
Solutions like $\phi_{K=2}$ obtained by solving the full equation in
eq.~(\ref{eq:11}) are therefore independent of the interactions and
the eigenvalue is independent of $\rho$.

\subsection{The \protect{$\lambda$}-spectrum for large $\rho$}

For large values of $\rho$ the short-range two-body potential $v$ with
range $b$ is non-vanishing only when $\alpha$ is smaller than a few
times $b/\rho$.  For larger values of $\alpha$ the ``rotation'' terms,
$\hat R\phi$, in the angular equation in eq.~(\ref{eq:35}) can then be
omitted.

For small values of $\alpha$ we find by substitution of
$r=\sqrt2\rho\sin\alpha$ instead of $\alpha$ in eq.~(\ref{eq:35}) that
the remaining equation (without rotation terms) can be rewritten as
\begin{eqnarray}
  0=\bigg(-\frac{\hbar^2}{2\mu}\frac{d^2}{dr^2}
  +V(r)-E\bigg)u(r) \; ,
  \label{e11}
\end{eqnarray}
where $2m\rho^2E/\hbar^2 = \lambda+9N/2-9$, $u(\sqrt2\rho\sin\alpha) =
\tilde\phi(\alpha)$, and the reduced mass $\mu=m/2$.

Each two-body bound state solution $E<0$ then corresponds to an
eigenvalue $\lambda$ diverging towards $-\infty$ as $-\rho^2$. Such
solutions do not produce significant rotation terms, which is
consistent with the omission in the derivation.  The structure of the
$N$-body system is given by the fully symmetrized wave function of two
particles in the bound state, while all other particles are far away
thus producing the large $\rho$.

All other solutions to eq.~(\ref{eq:35}) without bound two-body states
correspond to wave functions distributed over larger regions of
$\alpha$-space.  As the potentials then vanish for large $\rho$ we are
left with the free solutions, i.e.~the free spectrum of non-negative
$\lambda$-values is obtained in this limit of large $\rho$ in addition
to the discussed diverging eigenvalues for bound two-body states.

When the attractive potential contains precisely one bound two-body
state with zero energy, the eigenvalue $\lambda$ approaches a negative
constant for large $\rho$.  This can be understood by considering a
two-body state with energy slightly below zero, which forces $\lambda$
to diverge slowly as $-\rho^2$.  On the other hand, if the two-body
system is slightly unbound, $\lambda$ instead converges to zero, which
is the lowest eigenvalue of the free solutions.  Precisely at the
threshold it seems that $\lambda$ cannot decide and therefore remains
constant.  Thus for infinite two-body s-wave scattering length one
$\lambda$-value approaches a negative constant reached for large
$\rho$, when the average distance between particles is much larger
than the range of the interaction.

\section{Angular variational equation}
\label{sec:angul-vari-equat}

The angular equation is essential and solutions are not easily
obtained.  Proceeding with the Faddeev equations is one option, but we
prefer first to derive the optimal angular equation within the Hilbert
space defined by the form of the angular wave function in
eq.~(\ref{e77}). We also want to exploit that the two-body interaction
is of very short range compared with the size of the system.

\subsection{Variational approach}

The angular Schr\"odinger equation for fixed $\rho$ in eq.~(\ref{e26})
and the ansatz of the wave function in eq.~(\ref{e77}) allow us to
express the eigenvalue as an expectation value, i.e.
\begin{eqnarray}  \label{e67}
  \lambda
  =
  \frac{\langle\Phi|\hat h_\Omega|\Phi\rangle_\Omega}
  {\langle\Phi|\Phi\rangle_\Omega}
  =
  \frac{\langle\sum\phi_{ij}|\hat h_\Omega|\Phi\rangle_\Omega}
  {\langle\sum\phi_{ij}|\Phi\rangle_\Omega}
  \;.
\end{eqnarray}
Since $\hat h_\Omega$ is
invariant with respect to interchange of particles, the terms
$\langle\phi_{ij}|\hat h_\Omega|\Phi\rangle_\Omega=
\langle\phi_{kl}|\hat h_\Omega|\Phi\rangle_\Omega\nonumber$ are
identical and eq.~(\ref{e67}) simplifies to
\begin{eqnarray}
  \label{e36} 
  \lambda=
  \frac
  {\langle\phi_{12}|\hat h_\Omega|\sum\phi_{kl}\rangle_\Omega}
  {\langle\phi_{12}|\sum\phi_{kl}\rangle_\Omega}
  \;.
\end{eqnarray}
The total angular volume element is $d\Omega_{N-1} =
d\Omega_\alpha^{(N-1)}d\Omega_\eta^{(N-1)}d\Omega_{N-2}$, see
eq.~(\ref{e7}).  The integrands are independent of
$\Omega_\eta^{(N-1)}$ allowing to omit $d\Omega_\eta^{(N-1)}$ from the
integrations.  Using eq.~(\ref{e9}) we then obtain
\begin{eqnarray}
  \label{e38} 
  \int d\Omega_\alpha^{(N-1)}\;\phi_{12}^*\int d\Omega_{N-2}
  \big(\hat h_\Omega-\lambda\big)\sum_{k<l}^N\phi_{kl}
  =
  0
  \;.
\end{eqnarray}
The wave function component $\phi^*_{12}$ is now varied until the
lowest eigenvalue is obtained.  This gives the integro-differential
equation
\begin{eqnarray}
  \label{e40} 
  \int d\Omega_{N-2}
  \sum_{k<l}^N
  \bigg[
  (\hat\Lambda^2_{N-1}-\lambda)\phi_{kl}+
  v_{kl}\sum_{m<n}^N\phi_{mn}
  \bigg]
  =0
  \;,\quad
\end{eqnarray}
where the unknown functions, $\phi_{ij}=\phi(\alpha_{ij})$, in fact
all are the same identical functions of the different coordinates
$\alpha_{ij}$.  This result resembles the Faddeev-like equations of
eq.~(\ref{eq:11}).  Many terms are identical, e.g.~$\int
d\Omega_{N-2}\;v_{12}\phi_{34} = \int d\Omega_{N-2}
\;v_{12}\phi_{56}$, since particles $1$ and $2$ cannot distinguish
between other pairs of particles, see
appendix~\ref{sec:counting-terms} for the details.  Collecting all
terms yields
\begin{eqnarray}
  \label{eq:30}
  \int d\Omega_{N-2}\bigg[\Big(\hat\Pi^2_{12}+v_{12}-\lambda\Big)
  \phi_{12}+G(\tau,\alpha_{12})\bigg]
  =
  0
  \;,\;
\end{eqnarray}
where $\tau$ denotes angular coordinates apart from $\alpha_{12}$, and
\begin{eqnarray}
  \label{eq:28}
  &&
  G(\tau,\alpha_{12})
  =
  \frac12 n_2 \big(\hat\Pi^2_{34}
  +v(\alpha_{12})+v(\alpha_{34})-\lambda\big)\phi(\alpha_{34})
  \nonumber\\
  &&
  \quad
  +\frac12 n_2 v(\alpha_{34})\phi(\alpha_{12})+2n_1v(\alpha_{13})
  \big(\phi(\alpha_{12})+\phi(\alpha_{23})\big)
  \nonumber\\
  &&
  \quad
  +2n_1\big(\hat\Pi^2_{13}+v(\alpha_{12})+v(\alpha_{13})-\lambda\big)
  \phi(\alpha_{13})
  \nonumber\\
  &&
  \quad
  +n_3\Big(v(\alpha_{34})
  \big(\phi(\alpha_{35})+\phi(\alpha_{15})\big)+
  v(\alpha_{13})\phi(\alpha_{45})\Big)
  \nonumber\\
  &&
  \quad
  +2n_2v(\alpha_{13})\big(\phi(\alpha_{14})+\phi(\alpha_{24})+
  \phi(\alpha_{34})\big)
  \nonumber\\
  &&
  \quad
  +2n_2v(\alpha_{34})\phi(\alpha_{13})
  +\frac{1}{4} n_4 v(\alpha_{34})\phi(\alpha_{56})
  \;,
\end{eqnarray}
where $n_i=\prod_{j=1}^i(N-j-1)$, and $\hat\Pi_{ij}^2$ is defined from
eq.~(\ref{e18a}) with $k=N-1$ and with $\alpha_k$ replaced by
$\alpha_{ij}$.  In eq.~(\ref{eq:28}) all terms depend at most on
coordinates of the six particles $1$-$6$.  The first three terms in
eq.~(\ref{eq:30}) do not depend on the integration variables $\tau$
leaving only $G(\tau,\alpha_{12})$ for integration.

By appropriate choices \cite{smi77} of Jacobi systems, the relevant
degrees of freedom can be expressed in terms of the five vectors
$\vec\eta_{N-1},\ldots,\vec\eta_{N-5}$.  One is the argument of the
variational function and not an integration variable.  The remaining
twelve-dimensional integral is then evaluated with the corresponding
volume element $d\tau\propto \prod_{i=2}^{5} d\Omega_\alpha^{(N-i)}
d\Omega_\eta^{(N-i)}$ where the normalization is $\int d\tau=1$.  Then
eq.~(\ref{eq:30}) becomes
\begin{eqnarray}
  \label{e42} 
  \Big(\hat\Pi^2_{12}+v(\alpha_{12})-\lambda\Big)
  \phi(\alpha_{12})+\int d\tau\; G(\tau,\alpha_{12})
  = 0
  \;,\quad
\end{eqnarray}
where we used that the first terms are independent of the integration
variables.  Eq.~(\ref{e42}) is a linear integro-differential equation
in one variable containing up to five-dimensional integrals, see
appendix~\ref{sec:evaluation-terms}.

\subsection{Simple models for short-range interactions}

The short-range two-body interaction with s-wave scattering length
$a\sub s$ has in mean-field contexts \cite{dal99} been modelled by the
three-dimensional $\delta$-function potential
\begin{eqnarray}
  \label{eq:12}
  V(\vec r_{kl})
  =
  \frac{4\pi\hbar^2a\sub s}{m}\delta(\vec r_{kl})
  \;.
\end{eqnarray}

With the constant angular wave function, $\Phi_{K=0} =
\sum_{i<j}^N\phi_{K=0}(\alpha_{ij})$, the
expectation value of the angular Hamiltonian $\hat h_\Omega$ becomes
\begin{eqnarray}
  \label{eq:14}
  &&
  \lambda_{K=0}
  =
  \langle\Phi_{K=0}|\hat h_\Omega|\Phi_{K=0}\rangle_\Omega
  \nonumber\\
  &&
  =
  \Big\langle\Phi_{K=0}\Big|
  \sum_{k<l}^Nv_{kl}
  \Big|\Phi_{K=0}\Big\rangle_\Omega
  \;,
\end{eqnarray}
without contribution from angular kinetic energy.  With the
$\delta$-interaction in eq.~(\ref{eq:12}) we obtain (see also
\cite{boh98})
\begin{eqnarray}
  \label{eq:15}
  \lambda_{K=0}^\delta
  =
  \sqrt{\frac{2}{\pi}}
  \;
  \gammafktB{3N-3}{3N-6}
  \; N (N-1) \; \frac{a\sub s}{\rho}
  \;.
\end{eqnarray}

The $\delta$-interaction however does not contain the possibility of
studying the short-range properties such as bound two-body systems and
collapse of a condensate into clusters. An improvement is made by
using a finite-range, but still short-range, potential.  When $\rho$
is much larger than the potential range, $b$, we find an expectation
value of the angular energy of the same form as
$\lambda_{K=0}^\delta$ in eq.~(\ref{eq:15}):
\begin{eqnarray}
  \label{eq:16}
  \lambda_{K=0}\super{finite}
  \stackrel{\rho\gg b}\longrightarrow
  \sqrt{\frac{2}{\pi}}\;
  \gammafktB{3N-3}{3N-6}\;
  N (N-1)\;
  \frac{a\sub B}{\rho}
  \;.
\end{eqnarray}
The strength is now collected in the parameter $a\sub B$ instead of
$a\sub s$, where
\begin{eqnarray}
  \label{eq:17}
  a\sub B \equiv \frac{m}{4\pi\hbar^2}
  \int d^3\vec r_{kl}\; V(\vec r_{kl})
\end{eqnarray}
is the Born-approximation to the scattering length $a\sub s$.

However, in the opposite limit, when $\rho\ll b$, the result is
strongly dependent on the shape of the potential.  For example, for a
Gaussian potential $V(r_{kl}) = V_0\exp(-r_{kl}^2/b^2)$ with $a\sub
B=\sqrt\pi mb^3V_0/(4\hbar^2)$ we obtain
\begin{eqnarray}\label{eq:4}
  \lambda_{K=0}\super{finite}
  \stackrel{\rho\ll b}\longrightarrow
  \frac{4}{\sqrt\pi}N(N-1)\frac{a\sub B}{b}\Big(\frac{\rho}{b}\Big)^2
  \;.
\end{eqnarray}
As seen from these two limits there are some interesting
scaling-properties for finite-range potentials.  The angular
eigenvalue at a given $N$-value depends only on $a\sub B/b$ and
$\rho/b$.  More specifically we find for a Gaussian potential
\begin{eqnarray}
  v_{kl}
  =
  \frac{2m\rho^2V_0}{\hbar^2}e^{-r_{kl}^2/b^2}
  =
  \frac{8a\sub B}{\sqrt\pi b}
  \Big(\frac{\rho}{b}\Big)^2
  e^{-2(\rho/b)^2\sin^2\alpha_{kl}}
  \;,\qquad
\end{eqnarray}
which implies that for a given $a\sub B/b$ the angular eigenvalue
$\lambda$ is a function of $\rho/b$ only.  Moreover the radial
potential can be written as
\begin{eqnarray}
  \frac{2mb^2U}{\hbar^2}
  =
  \frac{\lambda}{(\rho/b)^2}+
  \frac{(3N-4)(3N-6)}{4(\rho/b)^2}+\frac{(\rho/b)^2}{(b\sub t/b)^4}
  \;,\quad
\end{eqnarray}
and the scaled energy, $2mb^2E/\hbar^2$, is then for a given $N$-value
a function of $a\sub B/b$ and $b\sub t/b$ only.  These scaling
properties are useful in model calculations.

\begin{figure}[htbp]
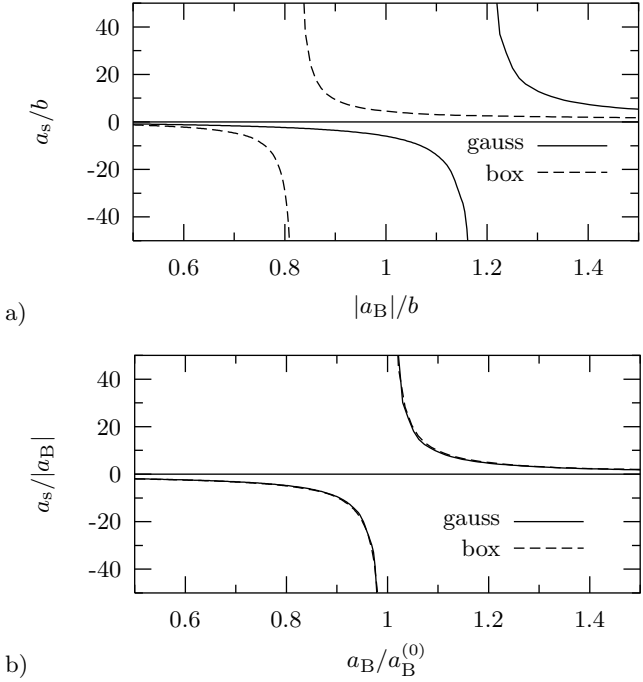

  \begin{center}
    %a)\input{/usr/users/oles/Few-body/Skriblerier/DelA/Scatteringlength/as_ap.tex}
    a)\input{fig4a.tex}
    %b)\input{/usr/users/oles/Few-body/Skriblerier/DelA/Scatteringlength/as_ap.scaled.tex}
    b)\input{fig4b.tex}
  \end{center}
  \vspace{-.5cm}
  \caption
  [] {a) Scattering length $a\sub s$ divided by the typical potential
    range $b$ as a function of $a\sub B$ divided by $b$.  b)
    Scattering length $a\sub s$ divided by $a\sub B$ as a function of
    $a\sub B$ divided by $a\sub B^{(0)}$ where the first bound state
    occurs.  Results are shown for the Gaussian potential
    $V(r)=V_0\exp(-r^2/b^2)$, ($a\sub B^{(0)}/b=-1.1893$), and for the
    box potential $V(r)=V_0\Theta(r<b)$, ($a\sub
    B^{(0)}/b=-0.82248$).}
  \label{fig4}
  \vspace{-.2cm}
\end{figure}

The scattering lengths are apparently essential parameters.  We show
in fig.~\ref{fig4}a the s-wave scattering length, $a\sub{s}$, as a
function of the strength parameter $|a\sub B|/b$ for both the Gaussian
potential and for an attractive box potential. When the numerical
value of the parameter $a\sub B$ is increased from zero, the
scattering length varies slowly and roughly linearly with $a\sub B$
for small $a\sub{B}$ until the value $a\sub{B}^{(0)}$, where
$a\sub{s}$ diverges as a signal of the appearance of the first
two-body bound state.  The threshold value of $a\sub B^{(0)}$ is
different for the Gaussian and for the box potentials, but
$a\sub{s}/a\sub B$ as a function of $a\sub B/ a\sub B^{(0)}$ results
in virtually the same curves, see fig.~\ref{fig4}b.  This indicates
that for simple potentials, the behaviour is approximately shape
independent.

\subsection{Short-range approximation}

The angular eigenvalue equation, eq.~(\ref{e42}), simplifies in the
limit when the two-body interaction range $b$ is much smaller than
$\rho$. Then the integrals are either analytical or reduce to
one-dimensional integrals. This substitution of a $\delta$-function is
only allowed for the potentials appearing under the integrals.
Otherwise the Thomas collapse occurs \cite{fed01}.  Thus apart from
the local terms containing $v(\alpha)$, the results only depend on
$a\sub B$ defined in eq.~(\ref{eq:17}).  For example the $\int
d\tau\;v(\alpha_{34})$-term reduces to
\begin{eqnarray}
  \int d\tau\; v(\alpha_{34})
  \simeq
  v_1(\alpha)
  \equiv
  2\sqrt{\frac{2}{\pi}}
  \;
  \gammafktB{3N-6}{3N-9}
  \frac{a\sub B}{\rho\cos^3\alpha}
  \;.\;
  \label{eq:18}
\end{eqnarray}
Similarly the $\int d\tau\;v(\alpha_{13})$-term reduces to
\begin{eqnarray}
  &&
  \int d\tau\; v(\alpha_{13})
  \simeq
  v_{2}(\alpha)
  \equiv
  \nonumber\\
  &&
  \frac{8}{3\sqrt3}\cos^{3N-11}\beta_0\;
  \Theta(\alpha<\pi/3)\;v_{1}(\alpha)
  \;,
  \label{sec:short-range-appr-2}
\end{eqnarray}
where $\sin\beta_0\equiv\tan\alpha/\sqrt3$ and $\Theta$ is the truth
function.  The remaining terms can in this limit be expressed through
$v_1(\alpha)$, $v_2(\alpha)$, $\hat R_{ij}^{(k)}$ from
eqs.~(\ref{eq:intf34}) and (\ref{eq:intf13}), and other related
rotation operators $\hat R_{ijkl}^{(n)}$.  Corresponding definitions
are given in appendix~\ref{sec:results-delta-limit}.

\begin{figure}[htbp]
  \begin{center}
    \setlength{\unitlength}{1.5mm}
    %a)\input{/usr/users/oles/Few-body/Skriblerier/Figurer/threebody.v13phi34.tex}
    %b)\input{/usr/users/oles/Few-body/Skriblerier/Figurer/threebody.delta13phi34.tex}
    a)
\begin{picture}(19,19.5)(0.8,-0.4)
\linethickness{1pt}
\thinlines
\put(2,7){\line(1,1){10}}
\put(2,7){\line(3,-1){15}}
\put(17,2){\line(0,1){10}}
\put(1,6){\makebox(0,0)[t]{$1$}}
\put(13,19){\makebox(0,0)[t]{$2$}}
\put(18,1){\makebox(0,0)[t]{$3$}}
\put(18,14){\makebox(0,0)[t]{$4$}}
\put(2,7){\circle*{1}}
\put(12,17){\circle*{1}}
\put(17,2){\circle*{1}}
\put(17,12){\circle*{1}}
\put(6,13){\makebox(0,0)[b]{$\phi_{12}^*$}}
\put(9,2){\makebox(0,0)[b]{$v_{13}$}}
\put(15,7){\makebox(0,0)[b]{$\phi_{34}$}}
\end{picture}
b)
\begin{picture}(18.9,16.3)(-0.6,3.2)
\linethickness{1pt}
\thinlines
\put(2,7){\line(1,1){10}}
\put(1,6){\makebox(0,0)[t]{$1,3$}}
\put(13,19){\makebox(0,0)[t]{$2$}}
\put(18,14){\makebox(0,0)[t]{$4$}}
\put(2,7){\circle*{1}}
\put(12,17){\circle*{1}}
\put(17,12){\circle*{1}}
\put(2,7){\line(3,1){15}}
\put(6,13){\makebox(0,0)[b]{$\phi_{12}^*$}}
\put(12,5){\makebox(0,0)[b]{$v_{13}\sim\delta(\vec r_{13})$}}
\put(15,8){\makebox(0,0)[b]{$\phi_{34}\sim\phi_{14}$}}
\end{picture}
  \end{center}
  \vspace{-.4cm}
  \caption
  [] {Simplifications due to short-range potentials.}
  \label{fig:variationsled.delta}
  \vspace{-.1cm}
\end{figure}
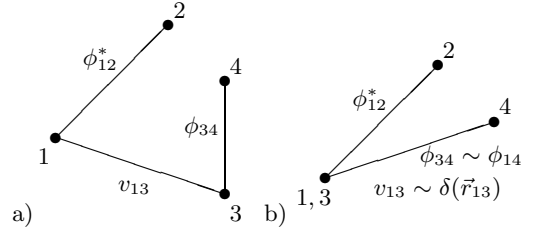

The reductions can be understood qualitatively via the diagrams in
fig.~\ref{fig:variationsled.delta}, which shows the geometry when the
zero-range interaction is contributing to the integrals.  As an
example, in the integral $\int
d\tau\;v(\alpha_{13})\phi(\alpha_{34})$, see
fig.~\ref{fig:variationsled.delta}a, particles $1$ and $3$ must be
close together as shown fig.~\ref{fig:variationsled.delta}b.  Then the
distance between particles $3$ and $4$ appearing in $\phi_{34}$ is
approximately equal to the distance between particles $1$ and $4$.
Therefore $\int d\tau\;v(\alpha_{13})\phi(\alpha_{34}) \simeq \int
d\tau\;v(\alpha_{13})\phi(\alpha_{14})$.
  
For $b \ll \rho$ we get almost independent of the potential
\begin{eqnarray}
  \label{e48}
  &&
  \int d\tau\; G(\tau,\alpha)
  \simeq
  \Big(\frac{n_2}{2}v_{1}(\alpha)
  +4n_1v_{2}(\alpha)\Big)\phi(\alpha)
  \nonumber\\
  &&
  \quad
  + \frac{n_2}{2} \hat R_{34}^{(N-2)}v\phi(\alpha)
  + 2n_1 \hat R_{13}^{(N-2)}v\phi(\alpha)
  \nonumber\\
  &&
  \quad
  +\frac{n_2}{2}\Big(\hat R_{34}^{(N-2)}\hat\Pi^2_{34}\phi(\alpha)+
  (v(\alpha)-\lambda)\hat R_{34}^{(N-2)}\phi(\alpha)\Big)
  \nonumber\\
  &&
  \quad
  +2n_1\Big(\hat R_{13}^{(N-2)}\hat\Pi^2_{13}\phi(\alpha)+
  (v(\alpha)-\lambda)\hat R_{13}^{(N-2)}\phi(\alpha)\Big)
  \nonumber\\
  &&
  \quad
  +\frac{n_4}4v_1(\alpha)\hat R_{34}^{(N-3)}\phi(\alpha)
  +n_3v_1(\alpha)\hat R_{3435}^{(1)}\phi(\alpha)
  \nonumber\\
  &&
  \quad
  +n_3v_2(\alpha)\hat R_{1345}^{(1)}\phi(\alpha)
  \nonumber\\
  &&
  \quad
  +n_3v_1(\alpha)\hat R_{13}^{(N-3)}\phi(\alpha)
  +2n_2v_1(\alpha)\hat R_{3413}^{(2)}\phi(\alpha)
  \nonumber\\
  &&
  \quad
  +2n_2v_2(\alpha)\big[2\hat R_{1314}^{(2)}\phi(\alpha)+
  \hat R_{1324}^{(2)}\phi(\alpha)\big]
  \;.
\end{eqnarray}
In the extreme zero-range limit the two terms $\hat R_{ij}^{(N-2)}
v\phi$ are proportional to $\phi(0)$. Using eqs.~(\ref{eq:18}) and
(\ref{sec:short-range-appr-2}) we define $\tilde v$ and $g$ by
\begin{equation}
  \label{eq:19}
  \tilde v(\alpha)
  \equiv
  \frac{n_2}{2}v_1(\alpha)
  \;\; , \;\;
  g(\alpha)\tilde v(\alpha) 
  \equiv 
  2 n_1v_2(\alpha)
\end{equation}
and rewrite eq.~(\ref{e48}) as
\begin{eqnarray}
  \label{eq:20}
  \int d\tau\; G(\tau,\alpha)
  &\simeq &
  \tilde v(\alpha)  
  \Big(1 + 2 g(\alpha) \Big){\phi(\alpha)} +
  \hat R \phi(\alpha)
  \nonumber\\
  &&+\tilde v(\alpha) \Big(1 + g(\alpha) \Big)\phi(0)\;,
\end{eqnarray}
where $\hat R\phi(\alpha)$ collectively denotes the remaining rotational
terms.  Collecting the terms from eq.~(\ref{e42}) then leads to an
integro-differential equation in one variable:
\begin{eqnarray}\label{eq:21}
  0
  &=&
  \Big(
  \hat\Pi^2_{12}
  +v(\alpha)
  +\tilde v(\alpha)\big(1+2g(\alpha)\big)
  -\lambda
  \Big)\phi(\alpha)
  \nonumber\\
  &&
  +\hat R\phi(\alpha)
  +\tilde v(\alpha) \big(1 + g(\alpha) \big)\phi(0)
  \;.
\end{eqnarray}
Eq.~(\ref{eq:21}) is a linear eigenvalue equation (in $\phi$) for one
variable $\alpha$. The eigenvalue $\lambda(\rho)$ is the key quantity
in the much simpler radial equation.

If we neglect the rotational terms in $\hat R$, assume $\phi(0)
\approx 0$, and use that $g(\alpha) \ll 1$ when $N$ is large, we get a
solution $\lambda \approx \tilde v(0)$.  Since $\tilde
v(0)=\lambda_{K=0}\super{finite}$ at large $N$, this eigenvalue is the
same as found in eq.~(\ref{eq:16}) from the expectation value of a
finite range potential in a constant angular wave function.
Eq.~(\ref{eq:21}) thus predicts an eigenvalue proportional to
$\rho^{-1}$ with a proportionality factor given by eq.~(\ref{eq:16}),
where the determining parameter is $a\sub B$.  However, as discussed
in \cite{boh98} the angular eigenvalue for small scattering length
$a\sub s$ is determined by $a\sub s$ only.  This is for a sufficiently
weak interaction the correct limit for $a\sub B$ as discussed in
connection with fig.~\ref{fig4}.

The repulsive effective potential in $\tilde v(\alpha)$ pushes the
wave function into a narrow region at small $\alpha$ outside the
repulsive core of $v(\alpha)$.  Such a solution is often not a good
approximation because the rotational terms are important.  Also for
attractive potentials the confinement to small $\alpha$ can not be
achieved by the $\tilde v(\alpha)$ term.  It is therefore crucial to
include all the terms of eq.~(\ref{eq:21}) as discussed in
\cite{sor01}.

\section{Numerical illustrations}
\label{sec:some-results}

The method has to be tested by solving the derived equations for
realistic parameter choices.  We first discuss qualitatively which
numerical technique we apply.  As we shall explain the difficulties
increase as $N$ and $\rho$ increase.  We shall therefore concentrate
on relatively small values of $N$, which happens to be a region of
growing interest in the art of making Bose-Einstein condensates. The
first physical results focus on the angular eigenvalues and the
related wave functions. Then the fully defined radial equation and its
solutions are discussed.

\subsection{Method of solving}

A usual method within the hyperspherical formalism is to expand the
angular wave function on kinetic energy eigenfunctions, the so-called
hyperspherical harmonics \cite{lin95,boh98}.  However, since the
hyperspherical harmonics contain oscillations at about $\alpha\sim
1/K$, large $K$'s of the order of $K\sub{max}\sim\rho/b$ are needed to
describe potentials limited to $\alpha < b/\rho$.  This $K\sub{max}$
becomes very large for the application to Bose-Einstein condensation,
so a basis of hyperspherical harmonics is not suitable in this
context.  Thus using only $K=0$ for a zero-range interaction
\cite{boh98} must be far from the optimal solution.

Instead we choose a basis of discrete mesh points distributed in
$\alpha$-space $\phi(\alpha)\to\underline\phi \equiv
[\phi(\alpha_1),\ldots, \phi(\alpha_m),\ldots,\phi(\alpha_M)]$ to take
into account the short range of the potential and to keep sufficient
information about small $\alpha$.  Derivatives are then written as
finite differences \cite{koo90} and integrations like $\hat
R\phi(\alpha)$ of eq.~(\ref{eq:21}) can be expressed in matrix form,
i.e.~$\hat R\phi(\alpha) \to \underline{\underline
  R}\;\underline\phi$.

Numerical computation of the integrals becomes increasingly difficult
with decreasing interaction range.  This can be understood in terms of
the $\alpha_{12}$ coordinate, since the potential at a given $\rho$
and a given range, $b$, of the interaction, is confined to an
$\alpha_{12}$-region of size $\Delta\alpha_{12}\sim b/\rho$, which for
Bose-Einstein condensates easily becomes very small and thus cannot be
handled directly numerically. In the following we shall use the
equations obtained in the short-range approximation, where the
difficulties are much smaller and the physics content is still
maintained.

\subsection{Angular solutions}

\begin{figure}[htb]
  \begin{center}
    \input{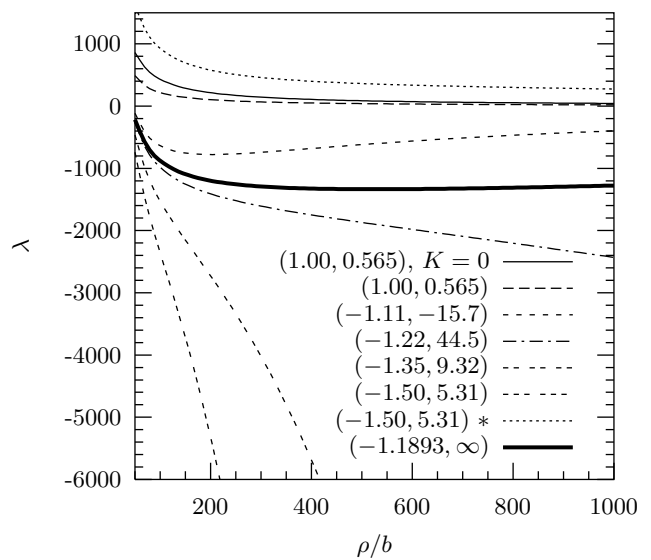}
  \end{center}
  \vspace{-.5cm}
  \caption 
  [] {Angular eigenvalues for $N=20$ and various parameters $(a\sub
    B/b,a\sub s/b)$ as shown on the figure.  A star refers to the
    first excited state.}
  \label{fig:lambdavar}
  \vspace{-.4cm}
\end{figure}

The behaviour of the lowest eigenvalues $\lambda$ depend strongly on
the potential. The characteristic feature is the large distance
asymptotic behaviour, i.e.~divergence as $-\rho^2$ corresponding to a
bound two-body state (see eq.~(\ref{e11})) or convergence as
$\rho^{-1}$ towards a finite value corresponding to the spectrum for
free particles.  The constant of proportionality to $\rho^{-1}$ is
qualitatively recovered as the predicted \cite{boh98} dependence on
$a\sub s$.  The exception arises for infinite scattering length at the
threshold for two-body binding, where one angular eigenvalue at large
$\rho$ approaches a negative constant.  This structure is illustrated
in fig.~\ref{fig:lambdavar}, where the lowest angular eigenvalue is
shown for various Gaussian potential strengths covering the region
from unbound to bound two-body states.

For repulsive potentials the eigenvalues are positive and the lowest
approaches zero from above. For weak attractions the lowest $\lambda$
is negative approaching zero from below. When the attraction can bind
a pair of particles the divergence is seen while the second $\lambda$
for the same potential is positive and approaching zero from above in
qualitative agreement with the lowest eigenvalue for a repulsive
potential. The higher eigenvalues would then converge to $K(K+3N-5)$
as $1/\rho$, where $K=4,6,8...$. The solution for $K=2$ is not
allowed, corresponding to removal of the non-physical spurious
solution of the Faddeev equations.  Increasing the attraction to allow
two bound two-body states would then shift the asymptotic spectrum
such that one more eigenvalue diverge while the non-negative energy
spectrum remains unchanged.

At each threshold for the appearance of a new bound two-body state one
eigenvalue asymptotically approaches a negative constant. This
especially interesting eigenvalue is responsible for the structure of
the $N$-body system for very large scattering lengths.

\begin{figure}[htb]
  \begin{center}
    \input{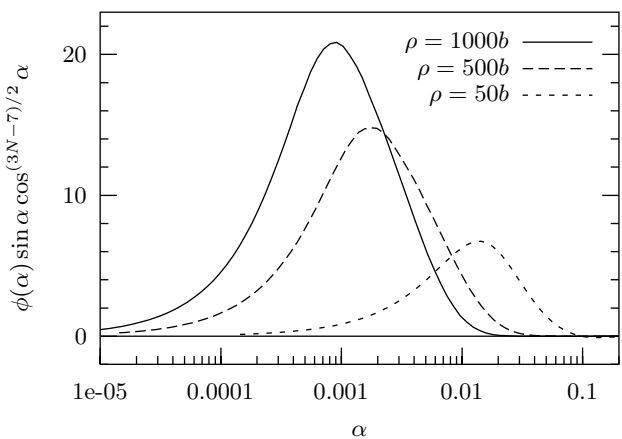}
  \end{center} 
  \vspace{-.6cm}
  \caption 
  [] {The lowest angular wave functions for $N=20$ and $a\sub B=-1.50b,
    a\sub s=5.31 b $ for three values of the hyperradius. This
    potential has one bound two-body state.}
  \label{fig:phibound}
  \vspace{-.3cm}
\end{figure}

The total angular wave functions are determined as a sum of the
components.  We show in fig.~\ref{fig:phibound} an example of the
reduced wave function for a potential with one bound two-body state.
The amplitude increases with $\rho$ and concentrates at very small
values of $\alpha$.  This reflects the convergence towards the
two-body bound state in agreement with the transformation $r_{12} =
\sqrt{2}\rho \sin \alpha$.  Recovering this behaviour numerically is
essential, otherwise the large distance behaviour cannot be described.

\begin{figure}[htb]
  \begin{center}
    \input{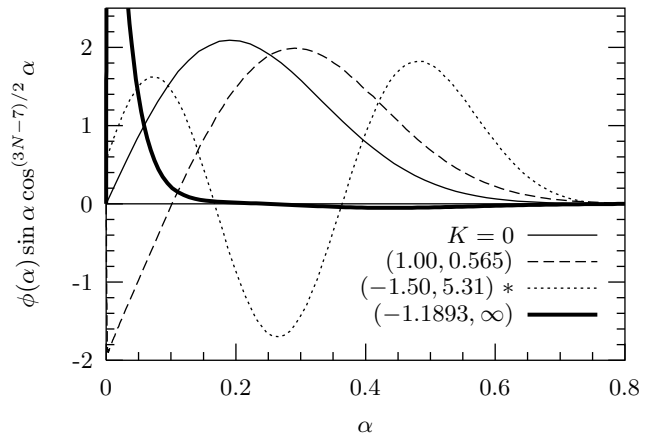}
  \end{center}
  \vspace{-.6cm}
  \caption 
  [] {Angular wave functions for $N=20$ and $\rho=500b$ for different
    interaction parameters $(a\sub B/b,a\sub s/b)$ as shown on the
    figure.  The $K=0$ curve corresponds to a non-interacting system.
    A star refers to the first excited state.}
  \label{fig:phi.asvar}
  \vspace{-.2cm}
\end{figure}

The angular eigenfunction varies with the strength of the interaction.
Examples of this variation are shown in fig.~\ref{fig:phi.asvar}.  The
non-interacting wave function has only nodes at the endpoints. The
repulsive case shows an oscillation, which lowers the angular energy
due to the rotation terms \cite{sor01}. The fast change at small
$\alpha$ is typical for interacting particles.  The wave function for
the excited state has an additional node. The corresponding
lower-lying wave function is shown in fig.~\ref{fig:phibound}.

The wave function for infinite scattering length corresponds to an
interaction where the two-body bound state is at the threshold for
occurrence.  This eigenfunction resembles those where a bound two-body
state is present, compare with the results shown in
fig.~\ref{fig:phibound} with different scales on both axes.  However,
now (thick curve of fig.~\ref{fig:phi.asvar}) the wave function is
located at larger $\alpha$-values and a node is present in the tail at
an intermediate $\alpha \sim 0.25$.

\begin{figure}[htb]
  \begin{center}
    \input{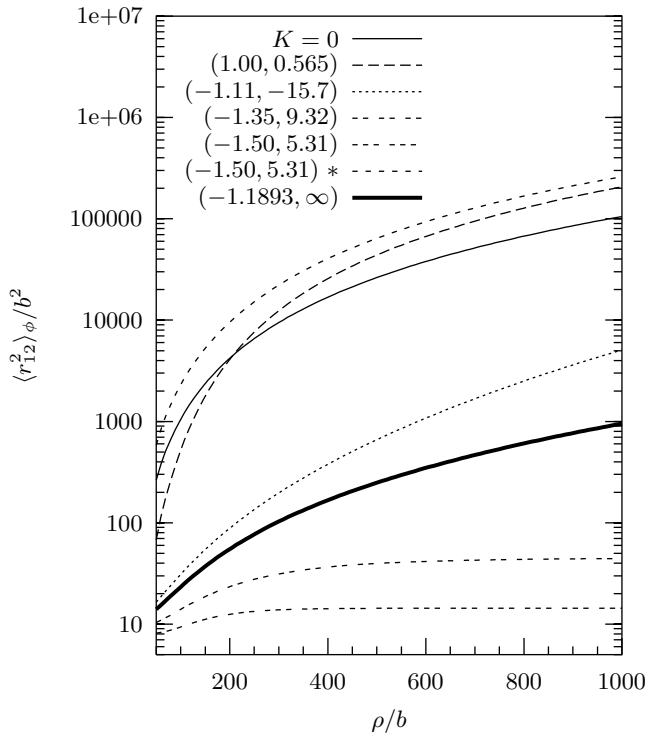}
  \end{center}
  \vspace{-.6cm}
  \caption 
  [] {The second moment $\langle r_{12}^2\rangle_\phi$ as a function
    of hyperradius for $N=20$ for solutions to the angular variational
    equation with different interaction parameters specified in the
    figure by $(a\sub B/b,a\sub s/b)$.  Also shown is the $K=0$-value.
    A star refers to the first excited state.}
  \label{fig:moment}
  \vspace{-.5cm}
\end{figure}

The structure of the component of the angular wave function is further
illustrated by the second moment defined by $\langle r_{12}^2
\rangle_{\phi} \equiv 2\rho^2\langle \phi | \sin^2\alpha |\phi
\rangle$. A number of these moments for different interactions are
shown in fig.~\ref{fig:moment} as functions of $\rho$.  For states
obtained from repulsive potentials, moderately attractive potentials
without bound two-body states, and for excited states of positive
$\lambda$, this moment, $\langle r_{12}^2\rangle_\phi$, increases
proportional to $\rho^2$ for large $\rho$.  This resembles the
behaviour of the expectation value in the lowest angular state for a
non-interacting system, $K=0$, where $\langle r_{12}^2\rangle_\phi =
2\rho^2/(N-1)$.  The qualitative explanation is that large $\rho$
implies the limit of a non-interacting spectrum with the corresponding
wave functions. The particles exploit the possibility of being far
from each other, since there is no energetic advantage of being close
due to the lack of a bound two-body state.

In contrast a different behaviour is observed when the potential can
bind two particles, i.e.~$\langle r_{12}^2\rangle_\phi$ approaches a
constant at large $\rho$.  This can be understood as the angular
equation in this limit approaches the two-body equation in
eq.~(\ref{e11}).  The radial wave function in the zero-range limit
converges to $u(r)=e^{-r/a\sub s}$, where $a\sub s$ is the scattering
length.  The second moment is then found as $\langle u|r^2|u\rangle =
a\sub s^2/2$, which in the limit of large $\rho$ reproduce $\langle
r_{12}^2\rangle_\phi$ for $a\sub s/b=9.32$ and $a\sub s/b=5.31$, see
fig.~\ref{fig:moment}.

This can be expressed in a different way: when a two-body bound state
is present, the angular wave function is at increasing $\rho$ squeezed
inside the potential, since the range in $\alpha$-space decreases
proportional to $\rho^{-1}$, and we obtain $\langle\phi| \sin^2\alpha
|\phi\rangle \propto 1/\rho^2$.  The distance between a pair of
particles is therefore independent of $\rho$ at large values of
$\rho$.  This means that pairwise the two-body bound state is
approached while all other particles are far away.  The symmetrization
does not affect this conclusion.

At the threshold for two-body binding, infinite scattering length, we
again observe the intermediate behaviour resembling a logarithmic
dependence in fig.~\ref{fig:moment}.

\subsection{Radial solutions}

\begin{figure}[htb]
  \begin{center}
    \input{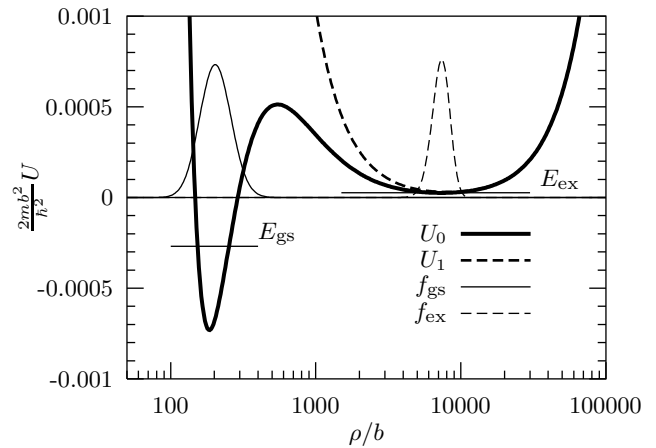}
  \end{center}
  \vspace{-0.6cm}
  \caption 
  [] {Radial potentials $U_0$ and $U_1$ from
    eq.~(\ref{eq:radial.potential}) corresponding to the two lowest
    angular potentials for $N=20$ and $a\sub s/b = -15.7$ in
    fig.~\ref{fig:lambdavar}.  We model the experimentally studied
    systems \cite{rob01} of $^{85}$Rb-atoms with oscillator frequency
    $\nu=\omega/(2\pi)=205$ Hz and interaction range $b=10$ a.u., thus
    yielding $b\sub t\equiv\sqrt{\hbar/(m\omega)}=1442b$.  Also shown
    are the two lowest energies and the radial eigenfunctions
    $f\sub{gs}$ and $f\sub{ex}$ for the lowest radial potential in the
    uncoupled radial equation, eq.~(\ref{e32}).}
  \label{fig:uasneg} 
  \vspace{-0.6cm}
\end{figure}

The most interesting angular eigenvalues either converge to zero or
remain constant for large $\rho$.  We therefore select an interaction
where $\lambda$ approaches zero relatively slowly from below. This is
then a weakly attractive potential without bound two-body states
although not very far from this threshold.  The parameter $a\sub s/b =
-15.7$ corresponds to this case. The resulting $\lambda$ shown in
fig.~\ref{fig:lambdavar} is used to compute the radial potential in
eq.~(\ref{eq:radial.potential}).  The total potential is shown in
fig.~\ref{fig:uasneg}.

The radial potential always diverges at large $\rho$ due to the
harmonic external field and at small $\rho$ due to the centrifugal
barrier term. Thus there is always infinitely many bound states.  For
the moderate attraction used in fig.~\ref{fig:uasneg} the potential
for the lowest angular eigenvalue has a global negative minimum at
small $\rho$ separated by a barrier at intermediate $\rho$ from a
local positive minimum at $\rho \sim \rho\sub
t\equiv\sqrt{3N/2}\;b\sub t$. The minimum at small $\rho$ disappears
quickly with a slightly less attractive potential. It becomes deeper
and wider for a more attractive potential, where the barrier at
intermediate $\rho$ simultaneously is reduced and eventually
disappears completely.

The radial potential corresponding to the second adiabatic angular
potential is also shown in fig.~\ref{fig:uasneg}. Since this $\lambda$
is positive for all $\rho$ an attractive pocket ($U<0$) in the radial
potential cannot arise. It coincides with the lowest radial potential
for large $\rho$ and due to the lack of attraction at small $\rho$ the
potential therefore diverges to $+ \infty$ for small $\rho$ without
going through another minimum.

The coupling terms of eq.~(\ref{eq:coupling.terms}) contribute at most
about 1 \% compared with other terms of the full radial equation,
eq.~(\ref{eq:coupled.radial.equation}), and can be omitted.  We
therefore arrive at solving the uncoupled radial equation,
eq.~(\ref{e32}), the solutions of which are considered in the
following.

The external field is negligible when $\rho\ll\sqrt{N} b\sub t$ and
the radial potential is therefore negative when
$\lambda+(3N-4)(3N-6)/4<0$ and $\rho$ is sufficiently small.  Then
genuinely bound many-body states of negative energy are possible in
our model without the influence of confinement from the trap.  The
radial equation corresponding to the parameters of the lowest
potential shown in fig.~\ref{fig:uasneg} has only one negative-energy
solution with the wave function located in the global negative
minimum.

The first of the infinitely many excited states in this potential is
located in the local minimum at larger $\rho$ created by competition
between the centrifugal barrier and the external harmonic oscillator
potential.  This excited state is usually referred to as the
Bose-Einstein condensate, because the corresponding wave function is
similar to that obtained in experiments \cite{dal99}.  It also
resembles the wave function found from the purely repulsive radial
potentials arising from both the second $\lambda$ with the present
attractive interaction and from repulsive two-body interactions.

From eq.~(\ref{e5}) we have $\langle\rho^2\rangle = N\langle
r_i^2\rangle -N\langle R^2\rangle = N\langle r_i^2\rangle-3b\sub
t^2/2$.  Using eq.~(\ref{e4}) we get $2 \langle\rho^2\rangle = (N-1)
\langle r_{12}^2\rangle$.  The root mean square radius $\langle
r_i^2\rangle^{1/2}$ of the condensate state, $f\sub{ex}$, is then 2 \%
larger and the energy $E\sub{ex}$ is 3 \% lower than the
Gross-Pitaevskii results.  The lowest radial state in $U_1$ has both
energy and root mean square radius about 10 \% larger than the
Gross-Pitaevskii results.

The moderate attraction used to obtain the potential in
fig.~\ref{fig:uasneg} is therefore seen to produce a lower-lying
many-body bound state with an average distance between the particles
about 37 times smaller than in the condensate state. The structure of
this state could as well be characterized as a condensate (condensed
$N$-body state), but it is much more unstable due to the many orders
of magnitude larger density and the subsequent larger recombination
probability \cite{nie99}.  This ground state, $f\sub{gs}$, has no
parallel in usual mean-field computations.

Even the state with $E\sub{gs}<0$, $\langle\rho^2\rangle^{1/2} \approx
216 b$, has a root mean square distance between two particles,
$\langle r_{12}^2\rangle^{1/2} \approx 70 b$, much larger than the
interaction range $b$.  This is a sufficiently large average distance
between particles to assume that the short range details of the
two-body potentials are unimportant.  Increasing the attraction more
negative-energy states appear in the attractive pocket inside the
external trap.  They occur at increasingly larger densities, but still
low enough for the short-range details to be unimportant, see more
details in \cite{sor02}.

At the two-body threshold, when the s-wave scattering length diverges,
potentially infinitely many negative-energy many-body states could
occur.  However, the external harmonic oscillator field limits the
number $N_E$ of these new negative-energy states, located inside the
trap and outside the two-body potential, to approximately
\begin{eqnarray}
  N_E \approx  0.5N\ln\bigg(\frac{ b\sub t} {37 b} \bigg) 
  \;,\quad
  \textrm{for }b\sub t\ll N|a\sub s|
  \;.
\end{eqnarray}  
The number of these multi-particle Efimov states therefore scales
linearly with $N$ and logarithmically with the ratio of trap length
and interaction range, see also \cite{sor02}.

The stability of the $N$-body system as a coherent object (condensate)
depends on the radial potential.  Starting with a condensate in the
minimum of $f\sub{ex}$ in fig.~\ref{fig:uasneg}, decay occurs both
through two- and three-body recombination processes and through
macroscopical tunneling through the barrier to state(s) in the global
minimum.  These denser states recombine faster due to the smaller
interparticle distances.  Increasing the attraction, the rate for
macroscopical tunneling increases due to a smaller barrier.
Eventually, at large (infinite) scattering length, the barrier has
vanished and the $N$-body system can contract freely, i.e.~populate
the many-body Efimov states.  The time scale for this contraction is
estimated to be about 0.1 ms \cite{sor02}, which can be compared with
experiments \cite{don01} and mean-field computations \cite{adh02b}.
Thus sudden removal of the barrier over the first short period of time
lead to spreading of the probability to the smaller distances where
recombination then takes place and the condensate decays with a
corresponding lifetime. This is qualitatively in complete agreement
with the measured decay function \cite{don01}.

\section{Summary and conclusions}

\label{sec:concl-persp}

We have formulated in details a new method to investigate two-body
correlations within symmetric $N$-body systems.  In contrast to the
product assumption for the wave function in the mean-field
approximation we use a Faddeev-type decomposition.  The overcomplete
basis is simplified by using in each of the two-body components only
the lowest possible number of partial waves, i.e.~s-waves. The allowed
Hilbert space is then dramatically reduced, it is not complete, but
hopefully precisely designed to treat the two-body correlations.

We use an adiabatic hyperspherical expansion with the hyperradius as
the adiabatic coordinate. We derive the optimal angular equation for
the two-body amplitude (the Faddeev component) arriving at an
one-dimensional angular integro-differential equation. Its eigenvalues
are closely related to the effective hyperradial potential, which
receives contributions from the two-body interactions, the kinetic
energy operator and from the external field confining the system. This
potential diverges both at small and large hyperradii due to the
centrifugal barrier and the external trap, respectively. 

For repulsive two-body interactions only one minimum in the radial
potential exists, but for moderately attractive two-body potentials
two minima are present separated by a barrier.  The minimum at
shortest hyperradii is at negative potential and is therefore able to
bind the system without help from the confining external field.  For a
state in this minimum the particles are on average still far outside
the range of interaction with each other.  The other minimum at larger
hyperradii can also be pronounced enough to hold localized stationary
states, which are similar to the solutions usually referred to as
Bose-Einstein condensates.

As the attraction increases and approaches the threshold for two-body
binding, i.e.~the scattering length increases towards infinity, the
intermediate barrier disappears and only the negative minimum at
smaller hyperradii remains.  However, it becomes both deeper and wider
and as a consequence more bound states of negative energy appear.
These many-body states have characteristic features of Efimov states.
They have inter-particle distances much larger than the range of the
two-body interaction and are therefore universal structures
independent of details of the potential.  Their number increases
proportional to the number of particles and logarithmically with the
ratio of trap length to interaction range.

In conclusion, we have formulated a method to treat two-body
correlations in an $N$-body system.  We applied the method to
Bose-Einstein condensates and obtained a simple one-dimensional
pictorial description in terms of an effective length coordinate.
Macroscopic collapse is conjectured to proceed via the new universal
Efimov states at intermediate hyperradii, which quickly, due to the
much larger density, subsequently recombine into dimer or trimer
states.  This decay can be studied quantitatively in the present
model.  Another unique feature of the model is to provide a solution
even in the case of very large scattering lengths, where the
Gross-Pitaevskii equation breaks.

\appendix

\section{Coordinates}

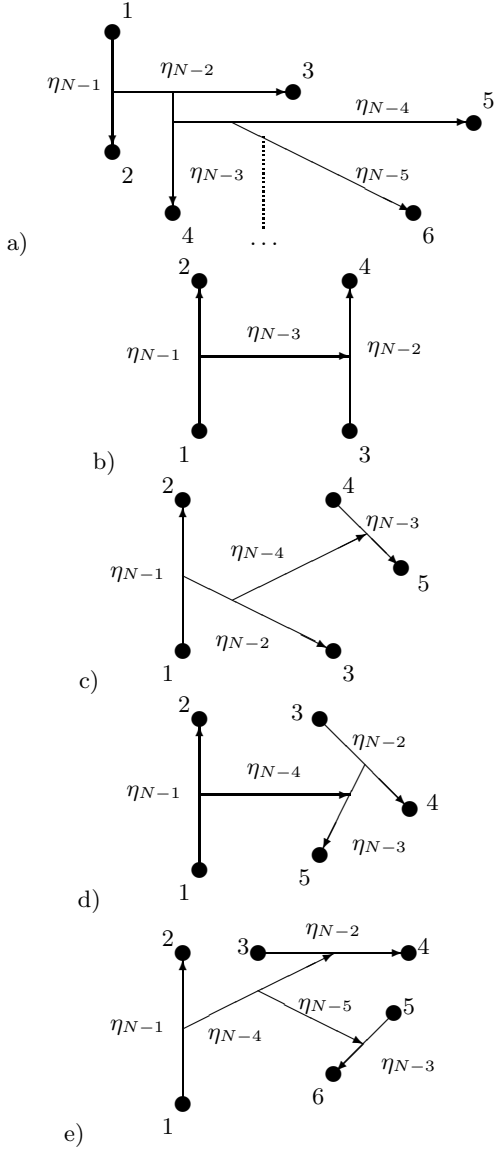
\begin{figure}[htb]
  \setlength{\unitlength}{1.7mm}
  \begin{center}
    %a)\input{/usr/users/oles/Few-body/Skriblerier/Figurer/sixplusbody.tex}
    %b)\input{/usr/users/oles/Few-body/Skriblerier/Figurer/phi12v34phi12.coord.tex}
    %c)\input{/usr/users/oles/Few-body/Skriblerier/Figurer/phi12v13phi45.coord.tex}
    %d)\input{/usr/users/oles/Few-body/Skriblerier/Figurer/phi12v34phi35.coord.tex}
    %e)\input{/usr/users/oles/Few-body/Skriblerier/Figurer/phi12v34phi56.coord.tex}
    a)
\setlength{\unitlength}{2mm}
\begin{picture}(30,16.5)(-4.5,-2.5)
\linethickness{1pt}
\thinlines
%\put(0,4){\vector(0,1){7.6}}
\put(0,12){\vector(0,-1){7.6}}
\put(0,8){\vector(1,0){11.6}}
\put(4,8){\vector(0,-1){7.6}}
\put(4,6){\vector(1,0){19.6}}
\put(8,6){\vector(2,-1){11.8}}
\put(1,14){\makebox(0,0)[t]{$1$}}
\put(1,3){\makebox(0,0)[t]{$2$}}
\put(13,10){\makebox(0,0)[t]{$3$}}
\put(5,-1){\makebox(0,0)[t]{$4$}}
\put(25,8){\makebox(0,0)[t]{$5$}}
\put(21,-1){\makebox(0,0)[t]{$6$}}
\put(0,4){\circle*{1}}
\put(0,12){\circle*{1}}
\put(12,8){\circle*{1}}
\put(4,0){\circle*{1}}
\put(24,6){\circle*{1}}
\put(20,0){\circle*{1}}
\put(-2.5,8){\makebox(0,0)[b]{$\eta_{N-1}$}}
\put(5,9){\makebox(0,0)[b]{$\eta_{N-2}$}}
\put(7,2){\makebox(0,0)[b]{$\eta_{N-3}$}}
\put(18,6.5){\makebox(0,0)[b]{$\eta_{N-4}$}}
\put(18,2){\makebox(0,0)[b]{$\eta_{N-5}$}}
\put(10.2,-2){\makebox(0,0)[t]{\ldots}}
\qbezier[20](10,5)(10,2)(10,-1)
\end{picture}
\\b)
\begin{picture}(19,14)(-5,-2.5)
\linethickness{1pt}
\thinlines
\put(0,0){\vector(0,1){9.6}}
\put(0,5){\vector(1,0){10}}
\put(10,0){\vector(0,1){9.6}}
\put(-1,-1){\makebox(0,0)[t]{$1$}}
\put(-1,11.5){\makebox(0,0)[t]{$2$}}
\put(11,-1){\makebox(0,0)[t]{$3$}}
\put(11,11.5){\makebox(0,0)[t]{$4$}}
\put(0,0){\circle*{1}}
\put(10,0){\circle*{1}}
\put(0,10){\circle*{1}}
\put(10,10){\circle*{1}}
\put(-3,4.6){\makebox(0,0)[b]{$\eta_{N-1}$}}
\put(13,5){\makebox(0,0)[b]{$\eta_{N-2}$}}
\put(5,6){\makebox(0,0)[b]{$\eta_{N-3}$}}
\end{picture}
\\c)
\begin{picture}(21,14)(-5,-2.5)
\linethickness{1pt}
\thinlines
\put(0,0){\vector(0,1){9.6}}
\put(0,5){\vector(2,-1){9.6}}
\put(3.33,3.33){\vector(2,1){8.9}}
\put(10,10){\vector(1,-1){4.3}}
\put(-1,-1){\makebox(0,0)[t]{$1$}}
\put(-1,11.5){\makebox(0,0)[t]{$2$}}
\put(11,-1){\makebox(0,0)[t]{$3$}}
\put(11,11.5){\makebox(0,0)[t]{$4$}}
\put(16,5){\makebox(0,0)[t]{$5$}}
\put(0,0){\circle*{1}}
\put(10,0){\circle*{1}}
\put(0,10){\circle*{1}}
\put(10,10){\circle*{1}}
\put(14.5,5.5){\circle*{1}}
\put(-3,4.6){\makebox(0,0)[b]{$\eta_{N-1}$}}
\put(4,0){\makebox(0,0)[b]{$\eta_{N-2}$}}
\put(14,8){\makebox(0,0)[b]{$\eta_{N-3}$}}
\put(5,6){\makebox(0,0)[b]{$\eta_{N-4}$}}
\end{picture}
\\d)
\begin{picture}(21,14)(-6,-2.5)
\linethickness{1pt}
\thinlines
\put(0,0){\vector(0,1){9.6}}
\put(0,5){\vector(1,0){10}}
\put(8,10){\vector(1,-1){5.7}}
\put(11,7){\vector(-1,-2){2.8}}
\put(-1,-1){\makebox(0,0)[t]{$1$}}
\put(-1,11.5){\makebox(0,0)[t]{$2$}}
\put(7,0){\makebox(0,0)[t]{$5$}}
\put(6.5,11){\makebox(0,0)[t]{$3$}}
\put(15.5,5){\makebox(0,0)[t]{$4$}}
\put(0,0){\circle*{1}}
\put(0,10){\circle*{1}}
\put(8,1){\circle*{1}}
\put(8,10){\circle*{1}}
\put(14,4){\circle*{1}}
\put(-3,4.6){\makebox(0,0)[b]{$\eta_{N-1}$}}
\put(12,8){\makebox(0,0)[b]{$\eta_{N-2}$}}
\put(12,1){\makebox(0,0)[b]{$\eta_{N-3}$}}
\put(5,6){\makebox(0,0)[b]{$\eta_{N-4}$}}
\end{picture}
\\e)
\begin{picture}(23,15)(-6,-2.5)
\linethickness{1pt}
\thinlines
\put(0,0){\vector(0,1){9.6}}
\put(0,5){\vector(2,1){10}}
\put(5,10){\vector(1,0){9.6}}
\put(5,7.5){\vector(2,-1){7}}
\put(14,6){\vector(-1,-1){3.8}}
\put(-1,-1){\makebox(0,0)[t]{$1$}}
\put(-1,11.5){\makebox(0,0)[t]{$2$}}
\put(4,11){\makebox(0,0)[t]{$3$}}
\put(16,11){\makebox(0,0)[t]{$4$}}
\put(15,7){\makebox(0,0)[t]{$5$}}
\put(9,1){\makebox(0,0)[t]{$6$}}
\put(0,0){\circle*{1}}
\put(0,10){\circle*{1}}
\put(5,10){\circle*{1}}
\put(15,10){\circle*{1}}
\put(14,6){\circle*{1}}
\put(10,2){\circle*{1}}
\put(-3,4.6){\makebox(0,0)[b]{$\eta_{N-1}$}}
\put(10,11){\makebox(0,0)[b]{$\eta_{N-2}$}}
\put(15,2){\makebox(0,0)[b]{$\eta_{N-3}$}}
\put(3.5,4){\makebox(0,0)[b]{$\eta_{N-4}$}}
\put(9.5,6){\makebox(0,0)[b]{$\eta_{N-5}$}}
\end{picture}
  \end{center}
  \vspace{-.5cm}
  \caption
  [] {Jacobi trees: a) standard, b) (12)(34), c) (123)(45), d)
    (12)(345), and e) (12)(34)(56).}
  \label{fig:jacobi.trees}
  \vspace{-.2cm}
\end{figure}

For use in the calculation of matrix elements different Jacobi trees
can be chosen \cite{smi77}.  The relevant ones in this context are
shown in fig.~\ref{fig:jacobi.trees}.

The coordinates of the standard tree of fig.~\ref{fig:jacobi.trees}a
are defined by
\begin{subequations}
  \label{eq:standard}
\begin{eqnarray}
  \vec\eta_{N-1}&&=
  \frac{1}{\sqrt2}(\vec r_2-\vec r_1)
  \;,
  \\
  \vec\eta_{N-2}&&=
  \sqrt{\frac{2}{3}}\Big(\vec r_3-\frac12(\vec r_2+\vec r_1)\Big)
  \;,
  \\
  \vec\eta_{N-3}&&=
  \sqrt{\frac{3}{4}}\Big(\vec r_4-\frac{1}{3}(\vec r_3+\vec r_2+\vec r_1)\Big)
  \;,
  \\
  &&\vdots
  \nonumber\\
  \vec\eta_1=&&
  \sqrt{\frac{N-1}{N}}\Big(\vec r_N-
  \frac{1}{N-1}(\vec r_{N-1}+\ldots+\vec r_1)\Big)
  \;.\qquad
\end{eqnarray}
\end{subequations}

In the (12)(34)-tree of fig.~\ref{fig:jacobi.trees}b two of the
vectors are different from the standard tree:
\begin{subequations}
  \label{eq:12.34}
\begin{eqnarray}
  \vec{\eta}_{N-2}&=&\frac1{\sqrt2}(\vec r_4-\vec r_3)
  \;,
  \\
  \vec{\eta}_{N-3}&=&\frac12(\vec r_4+\vec r_3-\vec r_2-\vec r_1)
  \;.
\end{eqnarray}
\end{subequations}

In the (123)(45)-tree of fig.~\ref{fig:jacobi.trees}c two of the
vectors differ from the standard tree:
\begin{subequations}
  \label{eq:123.45}
\begin{eqnarray}
  \vec{\eta}_{N-3}&=&\frac1{\sqrt2}(\vec r_5-\vec r_4)
  \;,
  \\
  \vec{\eta}_{N-4}&=&\sqrt{\frac65}\Big(\frac12(\vec r_5+\vec r_4)
  -\frac13(\vec r_3+\vec r_2+\vec r_1)\Big)
  \;.\;
\end{eqnarray}
\end{subequations}

In the (12)(345)-tree of fig.~\ref{fig:jacobi.trees}d three of the
vectors deviate from the standard tree:
\begin{subequations}
  \label{eq:12.345}
\begin{eqnarray}
  \vec{\eta}_{N-2}&=&\frac1{\sqrt2}(\vec r_4-\vec r_3)
  \;,
  \\
  \vec{\eta}_{N-3}&=&\sqrt{\frac23}(\vec r_5-\frac12(\vec r_4+\vec r_3))
  \;,
  \\
  \vec{\eta}_{N-4}&=&\sqrt{\frac65}\Big(\frac13(\vec r_5+\vec r_4+\vec r_3)
  -\frac12(\vec r_2+\vec r_1)\Big)
  \;.\;
\end{eqnarray}
\end{subequations}

In the (12)(34)(56)-tree of fig.~\ref{fig:jacobi.trees}e four of the
vectors are different from the standard tree:
\begin{subequations}
  \label{eq:12.34.56}
\begin{eqnarray}
  \vec{\eta}_{N-2}&=&\frac1{\sqrt2}(\vec r_4-\vec r_3)
  \;,
  \qquad
  \vec{\eta}_{N-3}=\frac1{\sqrt2}(\vec r_6-\vec r_5)
  \;,\qquad
  \\
  \vec{\eta}_{N-4}&=&\frac12(\vec r_4+\vec r_3-\vec r_2-\vec r_2)
  \;,
  \\
  \vec{\eta}_{N-5}&=&\sqrt{\frac43}\Big(\frac{\vec r_6+\vec r_5}2-
  \frac{\vec r_4+\vec r_3+\vec r_2+\vec r_1}4\Big)
  \;.\;
\end{eqnarray}
\end{subequations}

Since only inter-relations between $\vec\eta_{N-1}$, $\vec\eta_{N-2}$,
and $\vec\eta_{N-3}$ are needed in evaluating the matrix elements, we
will use the common notation:
\begin{eqnarray}
  \eta_{N-1}&=&\rho\sin\alpha
  \;,\qquad
  \eta_{N-2}=\rho\cos\alpha\sin\beta
  \\
  \eta_{N-3}&=&\rho\cos\alpha\cos\beta\sin\gamma
  \;,\quad
  \vec\eta_k\cdot\vec\eta_l=\eta_k\eta_l\cos\theta_{k,l}
  \;,\qquad
\end{eqnarray}
where $\theta_{k,l}$ is the angle between the $k$'th and $l$'th Jacobi
vectors.  We abbreviate $\theta_{N-1,N-2}\to\theta_x$,
$\theta_{N-1,N-3}\to\theta_y$, and $\theta_{N-2,N-3}\to\theta_z$.  The
azimuthal angle $\varphi$ determining the projection of
$\vec\eta_{N-3}$ onto the plane of $\vec\eta_{N-1}$ and
$\vec\eta_{N-2}$ is defined such that
\begin{eqnarray}
  \cos\theta_z&=&\sin\theta_x\sin\theta_y\cos\varphi+
  \cos\theta_x\cos\theta_y
  \;.
\end{eqnarray}
With $\tau=\{\beta,\gamma,\theta_x,\theta_y,\varphi\}$ a matrix
element of an arbitrary function $f$ of all the variables $\alpha$ and
$\tau$ then becomes
\begin{widetext}
\begin{eqnarray}
  \label{eq:6}
  \int d\tau\; f(\alpha,\tau)&=&
  \frac
  {\angleint{\beta}{3N-10}\angleint{\gamma}{3N-13}
    \int_0^\pi d\theta_x\sin\theta_x
    \int_0^\pi d\theta_y\sin\theta_y
    \int_0^{2\pi}d\varphi \;f(\alpha,\tau)}
  {\angleint{\beta}{3N-10}\angleint{\gamma}{3N-13}
    \int_0^\pi d\theta_x\sin\theta_x
    \int_0^\pi d\theta_y\sin\theta_y\int_0^{2\pi}d\varphi}
  \;,\;
\end{eqnarray}
\end{widetext}
where the normalization is $\int d\tau=1$.  We need relations for
interparticle distances and define $\vec\eta_{ij} \equiv (\vec
r_j-\vec r_i)/\sqrt2$ and the angle $\alpha_{ij}$ related to
$\eta_{ij} = \rho\sin\alpha_{ij}=r_{ij}/\sqrt2$.

\section{Integrals (Faddeev)}

\label{sec:app.fadd-like-equat}

Eqs.~(\ref{eq:intf34}) and (\ref{eq:intf13}) are evaluated as follows.

In the integral $\int d\tau\; \phi(\alpha_{34})$ a convenient choice
of coordinates is the alternative Jacobi (12)(34)-tree of
fig.~\ref{fig:jacobi.trees}b given by eqs.~(\ref{eq:12.34}).  The
angle $\alpha_{34}$ is associated with the distance
$r_{34}=\sqrt2\eta_{34}$ by the relation
\begin{eqnarray}
  \eta_{34}=
  \eta_{N-2}=
  \rho\cos\alpha\sin\beta=
  \rho\sin\alpha_{34}
  \\
  \Longleftrightarrow\;
  \sin\alpha_{34}=\cos\alpha\sin\beta
  \;.
  \label{eA1}
\end{eqnarray}
The integrand, $\phi(\alpha_{34})$, only depends on $\alpha_{34}$,
which is a function of $\alpha$ and $\beta$.  Therefore at fixed
$\alpha$ eq.~(\ref{eq:6}) reduces to
\begin{eqnarray}
  &&
  \int d\tau\; \phi(\alpha_{34})
  =
  \frac{\angleint{\beta}{3N-10}\;\phi(\alpha_{34})}
  {\int_0^{\pi/2}d\beta\;\sin^2\beta\cos^{3N-10}\beta}
  \nonumber\\
  &&=
  \frac{4}{\sqrt\pi}\gammafktB{3N-6}{3N-9}
  \int_0^{\pi/2}d\beta\;\sin^2\beta\cos^{3N-10}\beta\;\phi(\alpha_{34})
  \nonumber\\
  &&
  \equiv
  \hat R_{34}^{(N-2)}\phi(\alpha)
  \;,
  \label{eq:7}
\end{eqnarray}
where the last notation is convenient for repeated use.

\begin{figure}[htbp]
  \centering
  \input{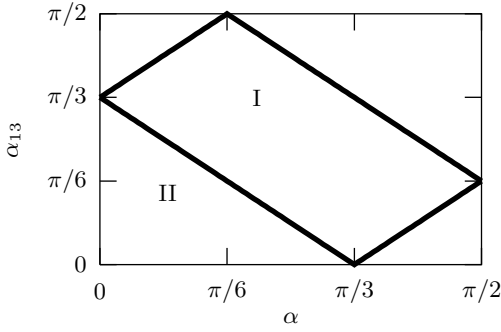}
  \vspace{-.2cm}
  \caption
  [] {The regions of integration in eq.~(\ref{eq:40}).}
  \label{fig:intarea}
  \vspace{-.2cm}
\end{figure}

To describe three particles in $\int d\tau\;\phi(\alpha_{13})$
simultaneously, Jacobi vectors of the standard tree are needed.  The
distance between particles $1$ and $3$ is related to the corresponding
Jacobi vector
\begin{eqnarray}
  \vec\eta_{13}=\frac{1}{\sqrt2}(\vec r_3-\vec r_1)=
  \frac12\vec\eta_{N-1}+\frac{\sqrt3}{2}\vec\eta_{N-2}
  \;,
\end{eqnarray}
The hyperangle $\alpha_{13}$, associated with the distance between
particles $1$ and $3$, through
$\eta_{13}=r_{13}/\sqrt2=\rho\sin\alpha_{13}$, is then
\begin{eqnarray}
  \label{eq:5}
  \sin^2\alpha_{13}
  =
  \frac{1}{4}\sin^2\alpha+
  \frac{3}{4}\cos^2\alpha\sin^2\beta+
  \nonumber\\
  \frac{\sqrt3}{2}\sin\alpha\cos\alpha\sin\beta\cos\theta_x
  \;,
\end{eqnarray}
where $\theta_x$ is the angle between the Jacobi vectors
$\vec\eta_{N-1}$ and $\vec\eta_{N-2}$.  Note that $\phi(\alpha_{13})$,
through $\alpha_{13}$, for fixed $\alpha$ depends on $\beta$ and
$\theta_x$, which leaves a two-dimensional integral. Therefore
eq.~(\ref{eq:6}) becomes
\begin{widetext}
\begin{eqnarray}
  \int d\tau\; \phi(\alpha_{13})
  &=&
  \frac{\int_0^{\pi/2}d\beta\;\sin^2\beta\cos^{3N-10}\beta
    \int_0^\pi d\theta_x\;\sin\theta_x\;\phi(\alpha_{13})}
  {\int_0^{\pi/2}d\beta\;\sin^2\beta\cos^{3N-10}\beta
    \int_0^\pi d\theta_x\;\sin\theta_x}
  \nonumber\\
  &=&
  \frac{2}{\sqrt\pi}\gammafktB{3N-6}{3N-9}
  \int_0^{\pi/2}d\beta\;\sin^2\beta\cos^{3N-10}\beta
  \int_0^\pi d\theta_x\;\sin\theta_x\;\phi(\alpha_{13})
  \;.
\end{eqnarray}
This integral can be reduced to one dimension by a partial
integration.  The final one-dimensional integral becomes
\begin{eqnarray}
  &&
  \int d\tau\;\phi(\alpha_{13})
  =
  \frac{4}{\sqrt{3\pi}}\gammafktB{3N-6}{3N-7}
  \sin^{-1}\alpha\cos^{8-3N}\alpha
  \Bigg[\int_{(\alpha-\pi/3)\Theta(\alpha>\pi/3)}^{\pi/2-|\pi/6-\alpha|}
  d\alpha_{13}\;\cos^{3N-9}\gamma^+\sin\alpha_{13}\cos\alpha_{13}
  \phi(\alpha_{13})
  \nonumber\\
  &&
  \qquad-\int_0^{(\pi/3-\alpha)\Theta(\pi/3>\alpha)}
  d\alpha_{13}\;\cos^{3N-9}\gamma^-\sin\alpha_{13}\cos\alpha_{13}
  \phi(\alpha_{13})\Bigg]
  \equiv\hat R_{13}^{(N-2)}\phi(\alpha)\;,
  \label{eq:40}
\end{eqnarray}
\end{widetext}
where $\sin^2\gamma^{\pm} = 4(\sin^2\alpha +\sin^2\alpha_{13}
\mp\sin\alpha\sin\alpha_{13})/3$, and $\Theta$ is the truth function.
The last notation is convenient for repeated use.  The integration
regions of the integrals are shown in fig.~\ref{fig:intarea}.  At
fixed $\alpha$ the range of $\alpha_{13}$ in the first integral is
over regions I and II, while in the second it is only over region II.
For $N=3$ the integrands of the two integrals are identical and thus
the integration over region II in the first integral cancels the
second integral.

\section{Integrals (variational)}

\label{sec:int.var.eqn}

We first divide the integrals of eq.~(\ref{e40}) into similar terms,
then compute them in general, and finally in the zero-range limit.

\subsection{Counting terms}

\label{sec:counting-terms}

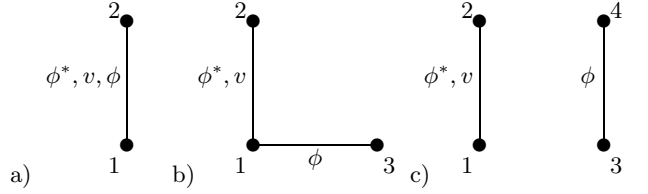
\begin{figure}[htbp]
  \setlength{\unitlength}{1.65mm}
  \begin{center}
    %a)\input{/usr/users/oles/Few-body/Skriblerier/Figurer/phi12v12phi12.tex}
    %b)\input{/usr/users/oles/Few-body/Skriblerier/Figurer/phi12v12phi13.tex}
    %c)\input{/usr/users/oles/Few-body/Skriblerier/Figurer/phi12v12phi34.tex}
    %a)\input{/usr/users/oles/Few-body/Skriblerier/Figurer/phi12v12phi12.tex}
%b)\input{/usr/users/oles/Few-body/Skriblerier/Figurer/phi12v12phi13.tex}
%c)\input{/usr/users/oles/Few-body/Skriblerier/Figurer/phi12v12phi34.tex}
a)
\begin{picture}(10,15)(-7,-3)
\linethickness{1pt}
\thinlines
\put(0,0){\line(0,1){10}}
\put(-1,-1){\makebox(0,0)[t]{$1$}}
\put(-1,11.5){\makebox(0,0)[t]{$2$}}
\put(0,0){\circle*{1}}
\put(0,10){\circle*{1}}
\put(-3.5,4.6){\makebox(0,0)[b]{$\phi^*,v,\phi$}}
\end{picture}
b)
\begin{picture}(16,15)(-4,-3)
\linethickness{1pt}
\thinlines
\put(0,0){\line(1,0){10}}
\put(0,0){\line(0,1){10}}
\put(-1,-1){\makebox(0,0)[t]{$1$}}
\put(-1,11.5){\makebox(0,0)[t]{$2$}}
\put(11,-1){\makebox(0,0)[t]{$3$}}
\put(0,0){\circle*{1}}
\put(0,10){\circle*{1}}
\put(10,0){\circle*{1}}
\put(-2.5,4.6){\makebox(0,0)[b]{$\phi^*,v$}}
\put(5,-2){\makebox(0,0)[b]{$\phi$}}
\end{picture}
c)
\begin{picture}(15,15)(-3.3,-3)
\linethickness{1pt}
\thinlines
\put(0,0){\line(0,1){10}}
\put(10,0){\line(0,1){10}}
\put(-1,-1){\makebox(0,0)[t]{$1$}}
\put(-1,11.5){\makebox(0,0)[t]{$2$}}
\put(11,-1){\makebox(0,0)[t]{$3$}}
\put(11,11.5){\makebox(0,0)[t]{$4$}}
\put(0,0){\circle*{1}}
\put(10,0){\circle*{1}}
\put(10,10){\circle*{1}}
\put(0,10){\circle*{1}}
\put(-2.5,4.6){\makebox(0,0)[b]{$\phi^*,v$}}
\put(8.7,4.6){\makebox(0,0)[b]{$\phi$}}
\end{picture}
  \end{center}
  \vspace{-.5cm}
  \caption
  [] {Illustration of $\phi_{12}^*v_{12}$-terms.}
  \label{fig:phi12v12}
  \vspace{-.2cm}
\end{figure}

We have to evaluate the double sums of eq.~(\ref{e40}) including the
potential:
\begin{eqnarray}
  \sum_{k<l}^Nv_{kl}
  \sum_{m<n}^N\phi_{mn}
  \;.
\end{eqnarray}
Three types of terms occur, due to the fact that we vary the wave
function component $\phi_{12}^*$ in eq.~(\ref{e38}): the potential
concerning particles $1$ and $2$, the potential concerning one of the
particles $1$ or $2$ and a third particle and the potential concerning
neither particle $1$ nor $2$, but a third and a fourth particle.  We
obtain
\begin{eqnarray}
  \sum_{k<l}^Nv_{kl}
  =
  v_{12}
  +\sum_{l=3}^Nv_{1l}
  +\sum_{l=3}^Nv_{2l}
  +\sum_{k\ge3,l>k}^Nv_{kl}
  \nonumber\\
  \to
  v_{12}+2(N-2)v_{13}+\frac12(N-2)(N-3)v_{34}
  \;,
\end{eqnarray}
where the arrow indicates the identity of the terms after integration
over all angles except $\alpha_{12}$ (analogously to the steps leading
up to eq.~(\ref{eq:angular-equation3})).  Treating each of these in
the quadruple sum, where the repeated use of arrows ($\to$) has the
meaning given just above:

Fixing $\phi_{12}^*$ and $v_{12}$ yields three different terms:
\begin{eqnarray}
  &&v_{12}
  \sum_{m<n}^N\phi_{mn}
  =
  \\
  &&v_{12}\Big(
  \phi_{12}
  +\sum_{n=3}^N\phi_{1n}
  +\sum_{n=3}^N\phi_{2n}
  +\sum_{m\ge3,n>m}^N\phi_{mn}
  \Big)
  \to
  \nonumber\\
  &&v_{12}
  \Big(
  \phi_{12}+2(N-2)\phi_{13}+\frac12(N-2)(N-3)\phi_{34}
  \Big)
  \nonumber
  \;,
\end{eqnarray}
as shown in fig.~\ref{fig:phi12v12}.

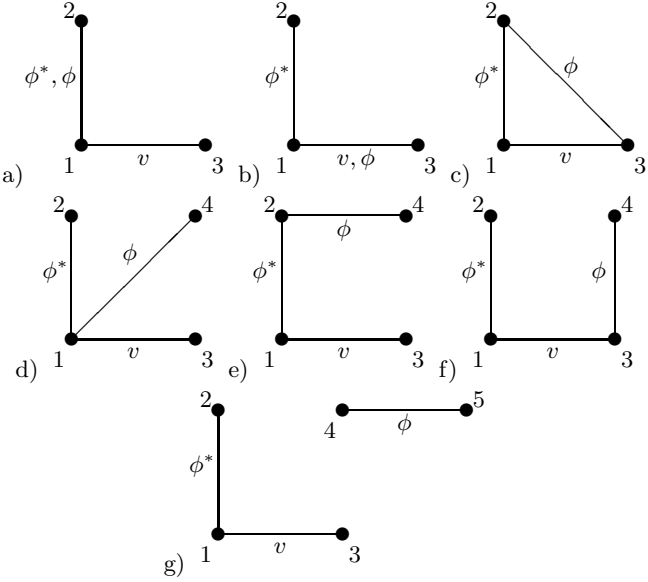
\begin{figure}[htbp]
  \setlength{\unitlength}{1.65mm}
  \begin{center}
%     a)\input{/usr/users/oles/Few-body/Skriblerier/Figurer/phi12v13phi12.tex}
%     b)\input{/usr/users/oles/Few-body/Skriblerier/Figurer/phi12v13phi13.tex}
%     c)\input{/usr/users/oles/Few-body/Skriblerier/Figurer/phi12v13phi23.tex}
%     d)\input{/usr/users/oles/Few-body/Skriblerier/Figurer/phi12v13phi14.tex}
%     e)\input{/usr/users/oles/Few-body/Skriblerier/Figurer/phi12v13phi24.tex}
%     f)\input{/usr/users/oles/Few-body/Skriblerier/Figurer/phi12v13phi34.tex}
%     g)\input{/usr/users/oles/Few-body/Skriblerier/Figurer/phi12v13phi45.tex}
    a)
\begin{picture}(16,15)(-4,-3)
\linethickness{1pt}
\thinlines
\put(0,0){\line(1,0){10}}
\put(0,0){\line(0,1){10}}
\put(-1,-1){\makebox(0,0)[t]{$1$}}
\put(-1,11.5){\makebox(0,0)[t]{$2$}}
\put(11,-1){\makebox(0,0)[t]{$3$}}
\put(0,0){\circle*{1}}
\put(0,10){\circle*{1}}
\put(10,0){\circle*{1}}
\put(-2.5,4.6){\makebox(0,0)[b]{$\phi^*,\phi$}}
\put(5,-1.6){\makebox(0,0)[b]{$v$}}
\end{picture}
b)
\begin{picture}(14,15)(-2,-3)
\linethickness{1pt}
\thinlines
\put(0,0){\line(1,0){10}}
\put(0,0){\line(0,1){10}}
\put(-1,-1){\makebox(0,0)[t]{$1$}}
\put(-1,11.5){\makebox(0,0)[t]{$2$}}
\put(11,-1){\makebox(0,0)[t]{$3$}}
\put(0,0){\circle*{1}}
\put(0,10){\circle*{1}}
\put(10,0){\circle*{1}}
\put(-1.3,4.6){\makebox(0,0)[b]{$\phi^*$}}
\put(5,-2){\makebox(0,0)[b]{$v,\phi$}}
\end{picture}
c)
\begin{picture}(14,15)(-2,-3)
\linethickness{1pt}
\thinlines
\put(0,0){\line(1,0){10}}
\put(0,0){\line(0,1){10}}
\put(0,10){\line(1,-1){10}}
\put(-1,-1){\makebox(0,0)[t]{$1$}}
\put(-1,11.5){\makebox(0,0)[t]{$2$}}
\put(11,-1){\makebox(0,0)[t]{$3$}}
\put(0,0){\circle*{1}}
\put(0,10){\circle*{1}}
\put(10,0){\circle*{1}}
\put(-1.3,4.6){\makebox(0,0)[b]{$\phi^*$}}
\put(5,-1.5){\makebox(0,0)[b]{$v$}}
\put(5.4,5.4){\makebox(0,0)[b]{$\phi$}}
\end{picture}
\\d)
\begin{picture}(14,15)(-2,-3)
\linethickness{1pt}
\thinlines
\put(0,0){\line(1,0){10}}
\put(0,0){\line(0,1){10}}
\put(0,0){\line(1,1){10}}
\put(-1,-1){\makebox(0,0)[t]{$1$}}
\put(-1,11.5){\makebox(0,0)[t]{$2$}}
\put(11,-1){\makebox(0,0)[t]{$3$}}
\put(11,11.5){\makebox(0,0)[t]{$4$}}
\put(0,0){\circle*{1}}
\put(10,0){\circle*{1}}
\put(10,10){\circle*{1}}
\put(0,10){\circle*{1}}
\put(-1.3,4.6){\makebox(0,0)[b]{$\phi^*$}}
\put(5,-1.5){\makebox(0,0)[b]{$v$}}
\put(4.7,6){\makebox(0,0)[b]{$\phi$}}
\end{picture}
e)
\begin{picture}(14,15)(-2,-3)
\linethickness{1pt}
\thinlines
\put(0,0){\line(1,0){10}}
\put(0,0){\line(0,1){10}}
\put(0,10){\line(1,0){10}}
\put(-1,-1){\makebox(0,0)[t]{$1$}}
\put(-1,11.5){\makebox(0,0)[t]{$2$}}
\put(11,-1){\makebox(0,0)[t]{$3$}}
\put(11,11.5){\makebox(0,0)[t]{$4$}}
\put(0,0){\circle*{1}}
\put(10,0){\circle*{1}}
\put(10,10){\circle*{1}}
\put(0,10){\circle*{1}}
\put(-1.3,4.6){\makebox(0,0)[b]{$\phi^*$}}
\put(5,-1.5){\makebox(0,0)[b]{$v$}}
\put(5,8){\makebox(0,0)[b]{$\phi$}}
\end{picture}
f)
\begin{picture}(14,15)(-2,-3)
\linethickness{1pt}
\thinlines
\put(0,0){\line(1,0){10}}
\put(0,0){\line(0,1){10}}
\put(10,0){\line(0,1){10}}
\put(-1,-1){\makebox(0,0)[t]{$1$}}
\put(-1,11.5){\makebox(0,0)[t]{$2$}}
\put(11,-1){\makebox(0,0)[t]{$3$}}
\put(11,11.5){\makebox(0,0)[t]{$4$}}
\put(0,0){\circle*{1}}
\put(10,0){\circle*{1}}
\put(10,10){\circle*{1}}
\put(0,10){\circle*{1}}
\put(-1.3,4.6){\makebox(0,0)[b]{$\phi^*$}}
\put(5,-1.5){\makebox(0,0)[b]{$v$}}
\put(8.7,4.6){\makebox(0,0)[b]{$\phi$}}
\end{picture}
\\g)
\begin{picture}(24,15)(-2,-3)
\linethickness{1pt}
\thinlines
\put(0,0){\line(1,0){10}}
\put(0,0){\line(0,1){10}}
\put(10,10){\line(1,0){10}}
\put(-1,-1){\makebox(0,0)[t]{$1$}}
\put(-1,11.5){\makebox(0,0)[t]{$2$}}
\put(11,-1){\makebox(0,0)[t]{$3$}}
\put(9,9){\makebox(0,0)[t]{$4$}}
\put(21,11.5){\makebox(0,0)[t]{$5$}}
\put(0,0){\circle*{1}}
\put(10,0){\circle*{1}}
\put(10,10){\circle*{1}}
\put(0,10){\circle*{1}}
\put(20,10){\circle*{1}}
\put(-1.3,4.6){\makebox(0,0)[b]{$\phi^*$}}
\put(5,-1.5){\makebox(0,0)[b]{$v$}}
\put(15,8){\makebox(0,0)[b]{$\phi$}}
\end{picture}
  \end{center}
  \vspace{-.5cm}
  \caption
  [] {Illustration of $\phi_{12}^*v_{13}$-terms.}
  \label{fig:phi12v13}
  \vspace{-.2cm}
\end{figure}

Fixing $\phi_{12}^*$ and $v_{13}$ yields seven different terms.  These
can be identified in two steps, the first of which separates into four
different sums:
\begin{eqnarray}
  &&v_{13}
  \sum_{m<n}^N\phi_{mn}
  =
  \\
  &&v_{13}
  \Big(
  \sum_{n=2}^N\phi_{1n}+
  \sum_{n=3}^N\phi_{2n}+
  \sum_{n=4}^N\phi_{3n}+
  \sum_{m\ge4,n>m}^N\phi_{mn}
  \Big)
  \nonumber
  \;.
\end{eqnarray}
Each of these four terms are then identified as:
\begin{eqnarray}
  v_{13}
  &&
  \sum_{n=2}^N\phi_{1n}
  =
  v_{13}\Big(\phi_{12}+\phi_{13}+\sum_{n=4}^N\phi_{1n}\Big)
  \nonumber\\
  &&
  \to
  v_{13}\big(\phi_{12}+\phi_{13}+(N-3)\phi_{14}\big)
  \;.
\end{eqnarray}
The similarity of the terms in the first sum becomes apparent when
carrying out the integration over $d\Omega_{N-2}$, e.g.~$\int
d\Omega_{N-2}v_{13}\phi_{14} = \int d\Omega_{N-2}v_{13}\phi_{15}$.
The rest are found in similar ways:
\begin{eqnarray}
  &&
  v_{13}
  \sum_{n=3}^N\phi_{2n}
  =
  v_{13}\Big(\phi_{23}+\sum_{n=4}^N\phi_{2n}\Big)
  \nonumber\\
  &&
  \qquad\to
  v_{13}\big(\phi_{23}+(N-3)\phi_{24}\big)
  \;,
  \\
  &&
  v_{13}
  \sum_{n=4}^N\phi_{3n}
  \to
  v_{13}(N-3)\phi_{34}
  \;,
  \\
  &&
  v_{13}
  \sum_{m\ge4,n>m}^N\phi_{mn}  
  \to
  v_{13}\frac12(N-3)(N-4)\phi_{45}
  \;.
\end{eqnarray}
The resulting seven types are shown in fig.~\ref{fig:phi12v13}.

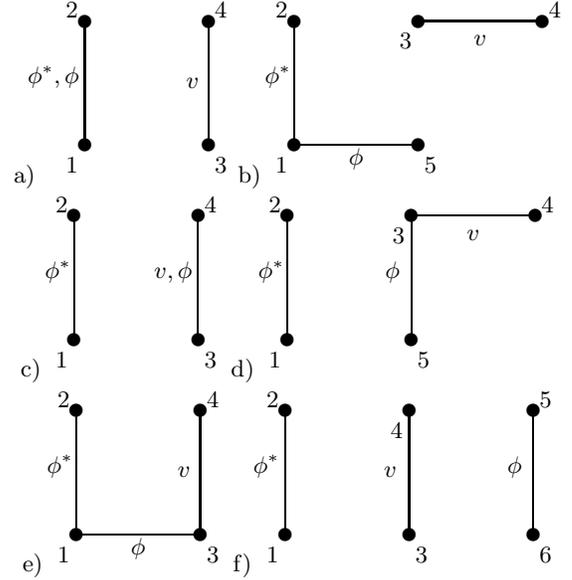
\begin{figure}[htbp]
  \setlength{\unitlength}{1.65mm}
  \begin{center}
%     a)\input{/usr/users/oles/Few-body/Skriblerier/Figurer/phi12v34phi12.tex}
%     b)\input{/usr/users/oles/Few-body/Skriblerier/Figurer/phi12v34phi15.tex}
%     c)\input{/usr/users/oles/Few-body/Skriblerier/Figurer/phi12v34phi34.tex}
%     d)\input{/usr/users/oles/Few-body/Skriblerier/Figurer/phi12v34phi35.tex}
%     e)\input{/usr/users/oles/Few-body/Skriblerier/Figurer/phi12v34phi13.tex}
%     f)\input{/usr/users/oles/Few-body/Skriblerier/Figurer/phi12v34phi56.tex}
    a)
\begin{picture}(15,15)(-3.3,-3)
\linethickness{1pt}
\thinlines
\put(0,0){\line(0,1){10}}
\put(10,0){\line(0,1){10}}
\put(-1,-1){\makebox(0,0)[t]{$1$}}
\put(-1,11.5){\makebox(0,0)[t]{$2$}}
\put(11,-1){\makebox(0,0)[t]{$3$}}
\put(11,11.5){\makebox(0,0)[t]{$4$}}
\put(0,0){\circle*{1}}
\put(10,0){\circle*{1}}
\put(10,10){\circle*{1}}
\put(0,10){\circle*{1}}
\put(-2.5,4.6){\makebox(0,0)[b]{$\phi^*,\phi$}}
\put(8.7,4.6){\makebox(0,0)[b]{$v$}}
\end{picture}
b)
\begin{picture}(24,15)(-2,-3)
\linethickness{1pt}
\thinlines
\put(0,0){\line(1,0){10}}
\put(0,0){\line(0,1){10}}
\put(10,10){\line(1,0){10}}
\put(-1,-1){\makebox(0,0)[t]{$1$}}
\put(-1,11.5){\makebox(0,0)[t]{$2$}}
\put(11,-1){\makebox(0,0)[t]{$5$}}
\put(9,9){\makebox(0,0)[t]{$3$}}
\put(21,11.5){\makebox(0,0)[t]{$4$}}
\put(0,0){\circle*{1}}
\put(10,0){\circle*{1}}
\put(10,10){\circle*{1}}
\put(0,10){\circle*{1}}
\put(20,10){\circle*{1}}
\put(-1.3,4.6){\makebox(0,0)[b]{$\phi^*$}}
\put(5,-2){\makebox(0,0)[b]{$\phi$}}
\put(15,8){\makebox(0,0)[b]{$v$}}
\end{picture}
\\c)
\begin{picture}(14,15)(-2,-3)
\linethickness{1pt}
\thinlines
\put(0,0){\line(0,1){10}}
\put(10,0){\line(0,1){10}}
\put(-1,-1){\makebox(0,0)[t]{$1$}}
\put(-1,11.5){\makebox(0,0)[t]{$2$}}
\put(11,-1){\makebox(0,0)[t]{$3$}}
\put(11,11.5){\makebox(0,0)[t]{$4$}}
\put(0,0){\circle*{1}}
\put(10,0){\circle*{1}}
\put(10,10){\circle*{1}}
\put(0,10){\circle*{1}}
\put(-1.3,4.6){\makebox(0,0)[b]{$\phi^*$}}
\put(8,4.6){\makebox(0,0)[b]{$v,\phi$}}
\end{picture}
d)
\begin{picture}(24,15)(-2,-3)
\linethickness{1pt}
\thinlines
\put(0,0){\line(0,1){10}}
\put(10,0){\line(0,1){10}}
\put(10,10){\line(1,0){10}}
\put(-1,-1){\makebox(0,0)[t]{$1$}}
\put(-1,11.5){\makebox(0,0)[t]{$2$}}
\put(11,-1){\makebox(0,0)[t]{$5$}}
\put(9,9){\makebox(0,0)[t]{$3$}}
\put(21,11.5){\makebox(0,0)[t]{$4$}}
\put(0,0){\circle*{1}}
\put(10,0){\circle*{1}}
\put(10,10){\circle*{1}}
\put(0,10){\circle*{1}}
\put(20,10){\circle*{1}}
\put(-1.3,4.6){\makebox(0,0)[b]{$\phi^*$}}
\put(8.5,4.6){\makebox(0,0)[b]{$\phi$}}
\put(15,8){\makebox(0,0)[b]{$v$}}
\end{picture}
\\e)
\begin{picture}(14,15)(-2,-3)
\linethickness{1pt}
\thinlines
\put(0,0){\line(1,0){10}}
\put(0,0){\line(0,1){10}}
\put(10,0){\line(0,1){10}}
\put(-1,-1){\makebox(0,0)[t]{$1$}}
\put(-1,11.5){\makebox(0,0)[t]{$2$}}
\put(11,-1){\makebox(0,0)[t]{$3$}}
\put(11,11.5){\makebox(0,0)[t]{$4$}}
\put(0,0){\circle*{1}}
\put(10,0){\circle*{1}}
\put(10,10){\circle*{1}}
\put(0,10){\circle*{1}}
\put(-1.3,4.6){\makebox(0,0)[b]{$\phi^*$}}
\put(5,-2){\makebox(0,0)[b]{$\phi$}}
\put(8.7,4.6){\makebox(0,0)[b]{$v$}}
\end{picture}
f)
\begin{picture}(24,15)(-2,-3)
\linethickness{1pt}
\thinlines
\put(0,0){\line(0,1){10}}
\put(10,0){\line(0,1){10}}
\put(20,0){\line(0,1){10}}
\put(-1,-1){\makebox(0,0)[t]{$1$}}
\put(-1,11.5){\makebox(0,0)[t]{$2$}}
\put(11,-1){\makebox(0,0)[t]{$3$}}
\put(9,9){\makebox(0,0)[t]{$4$}}
\put(21,11.5){\makebox(0,0)[t]{$5$}}
\put(21,-1){\makebox(0,0)[t]{$6$}}
\put(0,0){\circle*{1}}
\put(0,10){\circle*{1}}
\put(10,0){\circle*{1}}
\put(10,10){\circle*{1}}
\put(20,10){\circle*{1}}
\put(20,0){\circle*{1}}
\put(-1.5,4.6){\makebox(0,0)[b]{$\phi^*$}}
\put(8.5,4.6){\makebox(0,0)[b]{$v$}}
\put(18.5,4.6){\makebox(0,0)[b]{$\phi$}}
\end{picture}
  \end{center}
  \vspace{-.5cm}
  \caption
  [] {Illustration of $\phi_{12}^*v_{34}$-terms.}
  \label{fig:phi12v34}
  \vspace{-.2cm}
\end{figure}

Fixing $\phi_{12}^*$ and $v_{34}$ yields six different terms,
identified as follows.  The first step is:
\begin{eqnarray}
  &&v_{34}
  \sum_{m<n}^N\phi_{mn}
  =
  v_{34}
  \Big(
  \sum_{n=2}^N\phi_{1n}+
  \\
  &&
  \sum_{n=3}^N\phi_{2n}+
  \sum_{n=4}^N\phi_{3n}+
  \sum_{n=5}^N\phi_{4n}+
  \sum_{m\ge5,n>m}^N\phi_{mn}
  \Big)
  \nonumber
  \;.
\end{eqnarray}
In the next step the sums are treated:
\begin{eqnarray}
  &&
  v_{34}
  \sum_{n=2}^N\phi_{1n}
  =
  v_{34}\Big(
   \phi_{12}+\phi_{13}+\phi_{14}+\sum_{n=5}^N\phi_{1n}
  \Big)
  \nonumber
  %\\
\end{eqnarray}
\begin{eqnarray}
  &&
  \qquad\to
  v_{34}\big(\phi_{12}+2\phi_{13}+(N-4)\phi_{15}\big)
  \;,
  \\
  &&
  v_{34}
  \sum_{n=3}^N\phi_{2n}
  =
  v_{34}\Big(
   \phi_{23}+\phi_{24}+\sum_{n=5}^N\phi_{2n}
  \Big)
  \nonumber\\
  &&
  \qquad\to
  v_{34}\big(2\phi_{13}+(N-4)\phi_{15}\big)
  \;,
  \\
  &&
  v_{34}
  \sum_{n=4}^N\phi_{3n}
  =
  v_{34}\Big(
   \phi_{34}+\sum_{n=5}^N\phi_{3n}
  \Big)
  \nonumber\\
  &&
  \qquad\to
  v_{34}\big(\phi_{34}+(N-4)\phi_{35}\big)
  \;,
  \\
  &&
  v_{34}
  \sum_{n=5}^N\phi_{4n}
  \to
  v_{34}
  (N-4)\phi_{35}
  \;,
  \\
  &&
  v_{34}
  \sum_{m\ge5,n>m}^N\phi_{mn}
  \to
  v_{34}
  \frac12(N-4)(N-5)\phi_{56}
  \;.\quad
\end{eqnarray}
See the six types in fig.~\ref{fig:phi12v34}.

\subsection{Evaluation of terms}

\label{sec:evaluation-terms}

The term of fig.~\ref{fig:phi12v12}a is trivial since the integrand is
independent of $\tau$.  The terms of figs.~\ref{fig:phi12v12}b
\ref{fig:phi12v12}c, \ref{fig:phi12v13}a, \ref{fig:phi12v13}b,
\ref{fig:phi12v34}a, and \ref{fig:phi12v34}c can be evaluated by
eqs.~(\ref{eq:7}) and (\ref{eq:40}).

The term of fig.~\ref{fig:phi12v13}c becomes with the use of the
standard Jacobi tree of fig.~\ref{fig:jacobi.trees}a

\begin{widetext}

\begin{eqnarray}
  \int d\tau\; f(\alpha_{13})\;g(\alpha_{23})
  =
  \frac{2}{\sqrt\pi}\gammafktB{3N-6}{3N-9}
  \int_0^{\pi/2} d\beta\;\sin^2\beta\cos^{3N-10}\beta
  \int_0^\pi d\theta\;\sin\theta\;f(\alpha_{13})\;g(\alpha_{23})
  \;,
  %\\
\end{eqnarray}
\begin{eqnarray}
  \sin^2\alpha_{1(2)\;3}=\frac{3}{4}\cos^2\alpha\sin^2\beta+
  \frac{1}{4}\sin^2\alpha\pm
  \frac{\sqrt3}{2}\cos\alpha\sin\alpha\sin\beta\cos\theta
  \;.
\end{eqnarray}

The term of fig.~\ref{fig:phi12v34}f becomes with the use of the
alternative (12)(34)(56)-tree of fig.~\ref{fig:jacobi.trees}e
\begin{eqnarray}
  &&
  \int d\tau\; v(\alpha_{34})\;\phi(\alpha_{56})
  =
  \frac{2A_N}{\pi}
  \int d\beta\;\sin^2\beta\cos^{3N-10}\beta
  \int d\gamma\;\sin^2\gamma\cos^{3N-13}\gamma\;
  v(\alpha_{34})\;\phi(\alpha_{56})
  \;,
  \\
  &&
  \sin\alpha_{34}=\cos\alpha\sin\beta
  \;,\qquad
  \sin\alpha_{56}=\cos\alpha\cos\beta\sin\gamma
  \;,\qquad
  A_N\equiv (3N-8)(3N-10)(3N-12)
  \;.
\end{eqnarray}

The terms of figs.~\ref{fig:phi12v13}g, \ref{fig:phi12v34}b, and
\ref{fig:phi12v34}d are evaluated using the (123)(45)- and
(12)(345)-trees of figs.~\ref{fig:jacobi.trees}c and
\ref{fig:jacobi.trees}d, so
\begin{eqnarray}
  \int d\tau\; I_5(\alpha,\tau)
  =
  \frac{A_N}{\pi}
  \int_0^{\pi/2}d\beta\;\sin^2\beta\cos^{3N-10}\beta  
  \int_0^{\pi/2}d\gamma\;\sin^2\gamma\cos^{3N-13}\gamma
  \int_0^\pi d\theta_{x,z}\;\sin\theta_{x,z}\;
  I_5(\alpha,\tau)
  \;,
\end{eqnarray}
where $I_5(\alpha,\tau)$ can be either
$v(\alpha_{34})\phi(\alpha_{35})$ or $f(\alpha_{13})g(\alpha_{45})$.
The relevant angles are $\sin\alpha_{34}=\cos\alpha\sin\beta$,
$\sin^2\alpha_{35}=\cos^2\alpha(3\cos^2\beta\sin^2\gamma+
\sin^2\beta+2\sqrt3\cos\beta\sin\beta\sin\gamma\cos\theta_z)/4$,
$\sin\alpha_{45}=\cos\alpha\cos\beta\sin\gamma$, and $\alpha_{13}$ 
given by eq.~(\ref{eq:5}).  Note that the integral $\int d\tau\;
v(\alpha_{34})\phi(\alpha_{15})= \int d\tau\;
\phi(\alpha_{13})\;v(\alpha_{45})$.

The terms of figs.~\ref{fig:phi12v13}d, \ref{fig:phi12v13}e,
\ref{fig:phi12v13}f, and \ref{fig:phi12v34}e are evaluated using the
standard Jacobi tree.  Then eq.~(\ref{eq:6}) reduces to
\begin{eqnarray}
  \int d\tau\; f(\alpha_{13})\;g(\alpha_{i4})
  &=&
  \frac{A_N}{4\pi^2}
  \int_0^{\pi/2}d\beta\;\sin^2\beta\cos^{3N-10}\beta
  \int_0^{\pi/2}d\gamma\;\sin^2\gamma\cos^{3N-13}\gamma
  \nonumber\\
  &&
  \int_0^\pi d\theta_x\;\sin\theta_x
  \int_0^\pi d\theta_y\;\sin\theta_y
  \int_0^{2\pi} d\varphi\;
  f(\alpha_{13})\;g(\alpha_{i4})
  \;;\qquad i=1,2,3\;.
\end{eqnarray}
The angles $\alpha_{ij}$ can be determined by $\rho\sin\alpha_{ij} =
\eta_{ij}$ through the relations
\begin{eqnarray}
  &&\vec\eta_{13}
  =
  \frac{\sqrt3}{2}\vec\eta_{N-2}+\frac12\vec\eta_{N-1}
  \;,\qquad
  \vec\eta_{14}=\sqrt{\frac{2}{3}}\vec\eta_{N-3}+
  \frac{1}{2\sqrt3}\vec\eta_{N-2}+\frac12\vec\eta_{N-1}\;,
  \\
  &&\vec\eta_{24}=\sqrt{\frac{2}{3}}\vec\eta_{N-3}+
  \frac{1}{2\sqrt3}\vec\eta_{N-2}-\frac12\vec\eta_{N-1}\;,\qquad
  \vec\eta_{34}=\sqrt{\frac{2}{3}}\vec\eta_{N-3}-
  \frac{1}{\sqrt3}\vec\eta_{N-2}\;.
\end{eqnarray}
\end{widetext}

\subsection{Results in the $\delta$-limit}

\label{sec:results-delta-limit}

The integrals in the short-range limit, when the range $b$ of
$V(r_{ij})$ is much smaller than the size scale $\rho$, are:
\begin{eqnarray}
  &&
  \int d\tau\; v(\alpha_{34})\phi(\alpha_{13})\;\simeq\;
  v_1(\alpha)\hat R_{3413}\super{(2)}\phi(\alpha)
  \;,
  \\
  &&
  \int d\tau\; v(\alpha_{34})\phi(\alpha_{15})\;\simeq\;
  v_1(\alpha)\hat R_{13}^{(N-3)}\phi(\alpha)
  \;,
  \\
  &&
  \int d\tau\; v(\alpha_{34})\phi(\alpha_{34})=
  \hat R_{34}^{(N-2)}v\phi(\alpha)
  \simeq v_1(\alpha)\phi(0)
  \;,\;\qquad
  \\
  &&
  \int d\tau\; v(\alpha_{34})\phi(\alpha_{35})\;\simeq\;
  v_1(\alpha)\hat R_{3435}\super{(1)}\phi(\alpha)
  \;,
  \\
  &&
  \int d\tau\; v(\alpha_{34})\phi(\alpha_{56})\;\simeq\;
  v_1(\alpha)\hat R_{34}^{(N-3)}\phi(\alpha)
  \;,
\end{eqnarray}
\begin{eqnarray}
  &&
  \int d\tau\; v(\alpha_{13})\phi(\alpha_{13})=
  \hat R_{13}^{(N-2)}v\phi(\alpha)
  \simeq v_2(\alpha)\phi(0)
  \;,\qquad
  \\
  &&
  \int d\tau\; v(\alpha_{13})\phi(\alpha_{i4})\simeq
  v_2(\alpha)\hat R_{1314}\super{(2)}\phi(\alpha)
  \;;\; i=1,3
  \;,\quad
  \\
  &&
  \int d\tau\; v(\alpha_{13})\phi(\alpha_{23})\simeq
  v_2(\alpha)\phi(\alpha)
  \;,
  \\
  &&
  \int d\tau\; v(\alpha_{13})\phi(\alpha_{24})\simeq
  v_2(\alpha)\hat R_{1324}\super{(2)}\phi(\alpha)
  \;,
  \\
  &&
  \int d\tau\; v(\alpha_{13})\phi(\alpha_{45})\simeq
  v_2(\alpha)\hat R_{1345}\super{(1)}\phi(\alpha)
  \;.
\end{eqnarray}
The integrals are given by
\begin{widetext}
\begin{eqnarray}
  %&&
  \hat R_{ijkl}\super{(1)}\phi(\alpha)
  \equiv
  %\nonumber\\
  %&&
  \frac4{\sqrt\pi}\gammafktB{3N-9}{3N-12}\angleint{\gamma}{3N-13}
  \;\phi(\alpha_{kl}^0)
  \;,\qquad
\end{eqnarray}
where $\sin\alpha_{35}^0\equiv\sqrt3\cos\alpha\sin\gamma/2$,
$\sin\alpha_{45}^0\equiv\cos\alpha\cos\beta_0\sin\gamma$,
$\sin\beta_0\equiv\tan\alpha/\sqrt3$, and
\begin{eqnarray}
  &&
  \hat R_{ijkl}\super{(2)}\phi(\alpha)
  \equiv
  \frac2{\sqrt\pi}\gammafktB{3N-9}{3N-12}\angleint{\gamma}{3N-13}\int_0^\pi
  d\theta_x\sin\theta_x\;\phi(\alpha_{kl}^0)
  \;,
  \\
  &&
  \sin^2\alpha_{14}^0
  \equiv
  \frac19\sin^2\alpha+
  \frac23\cos^2\alpha\cos^2\beta_0\sin^2\gamma+
  \frac{2\sqrt2}{3\sqrt3}\sin\alpha\cos\alpha\cos\beta_0\sin\gamma\cos\theta_x
  \;,
  \\
  &&
  \sin^2\alpha_{24}^0
  \equiv
  \frac49\sin^2\alpha+
  \frac23\cos^2\alpha\cos^2\beta_0\sin^2\gamma+
  \frac{4\sqrt2}{3\sqrt3}\sin\alpha\cos\alpha\cos\beta_0\sin\gamma\cos\theta_x
  \;,
  \\
  &&
  \sin^2\alpha_{13}^0
  \equiv
  \frac14\sin^2\alpha+
  \frac12\cos^2\alpha\sin^2\gamma+
  \frac1{\sqrt2}\sin\alpha\cos\alpha\sin\gamma\cos\theta_x
  \;.
\end{eqnarray}
\end{widetext}
$\hat R_{ijkl}\super{(2)}\phi(\alpha)$ appears as a two-dimensional
integral but can be reduced to a one-dimensional integral, analogously
to eq.~(\ref{eq:40}).

%\bibliographystyle{prsty}
%\bibliography{/usr/users/oles/Few-body/Skriblerier/bibliografi.bib}

\end{document}